\documentclass{article}

\usepackage[english]{babel}
\usepackage{amsmath}
\usepackage[linesnumbered,ruled]{algorithm2e}

\usepackage[letterpaper,top=2cm,bottom=2cm,left=3cm,right=3cm,marginparwidth=1.75cm]{geometry}

\usepackage{amsmath}
\usepackage{graphicx}
\usepackage{subcaption}

\usepackage{tabularx}

\usepackage[colorlinks=true, allcolors=blue]{hyperref}
\usepackage[capitalise]{cleveref}

\DeclareMathOperator{\erf}{erf}
\DeclareMathOperator{\ierf}{ierf}
\DeclareMathOperator{\erfi}{erfi}
\DeclareMathOperator{\lerp}{lerp}

\makeatletter
\DeclareRobustCommand\onedot{\futurelet\@let@token\@onedot}
\def\@onedot{\ifx\@let@token.\else.\null\fi\xspace}
\def\etal{\emph{et al}\onedot}
\makeatother

\title{DiffTetVR: Differentiable Tetrahedral Volume Rendering}
\author{Christoph Neuhauser}

\begin{document}
\maketitle


Differentiable rendering is a technique that aims to invert the rendering process to enable optimizing rendering parameters from a set of images. In this article, we present a differentiable volume rendering solution called DiffTetVR for tetrahedral meshes. Unlike previous works based on regular grids, this enables the optimization of vertex positions and the local subdivision of the mesh without relying on multigrid methods. We present an efficient implementation of the forward rendering process, deduce the derivatives for the backwards pass and regularization terms for avoiding degenerate tetrahedra, and finally show how the tetrahedral mesh can be subdivided locally to enable a coarse-to-fine optimization process. The source code is made publicly available on GitHub at \url{https://github.com/chrismile/DiffTetVR}.\\\newline
\textbf{Disclaimer:} While the method presented in this article is promising, I have chosen to not further pursue it, as I have left academia after finishing my PhD thesis. The results section summarizes the intermediate results and weaknesses of the technique that need to be overcome to reach (and potentially outperform) state-of-the-art results. I encourage readers to take inspiration from the presented ideas and to reuse code that was written as part of this work to build future works that are able to achieve the goal DiffTetVR set out to solve and enable adaptive reconstruction of volumetric data from images.

\section{Introduction}

Differentiable rendering~\cite{DiffRenderSurvey} is a technique that has gained attention in the computer graphics and vision communities in recent years. While the forward rendering process is concerned with generating images from a set of rendering parameters such as geometry, materials and textures, differentiable rendering aims to invert this process in order to optimize the rendering parameters to best match a set of images. Differentiable rendering has been used in previous works for reversing the rendering process for representations such as meshes~\cite{SoftRas,NvDiffRast}, point clouds~\cite{LinPointCloud} and 3D Gaussian splats~\cite{Kerbl3DGS}. A recent survey by Kato~\etal~\cite{DiffRenderSurvey} gives an overview of this wide field of works.

Differentiable direct volume rendering (DiffDVR) was first introduced by Weiss and Westermann~\cite{DiffDVR}. Their work aims to invert the direct volume rendering process, which is concerned with generating images for volumetric fields. This rendering process usually involves marching through the volume along view rays and using alpha blending operations~\cite{PorterDuff} for blending semi-transparent colors sampled along the ray. Weiss and Westermann~\cite{DiffDVR} use DiffDVR for solving tasks such as automatic viewpoint selection, or transfer function, density field and color field reconstruction. Unlike previous works such as the Mitsuba 2 rendering framework \cite{Mitsuba2}, Weiss and Westermann do not require intermediate values to be saved for backpropagation, which strongly reduces the memory requirements of their method. Kerbl~\etal~\cite{Kerbl3DGS} later chose a similar strategy for inverting the alpha blending process in their differentiable 3D Gaussian splat (3DGS) optimizer, which has gained immense popularity in the field of differentiable rendering.

A limitation of the approach by Weiss and Westermann~\cite{DiffDVR} is that they are constrained to regular grids. This makes it harder to not only optimize grid properties such as stored densities or colors, but also the grid vertex positions. Furthermore, local subdivision of the mesh during the training process without relying on multigrid methods becomes infeasible. Thus, in this work, we introduce an approach for the differentiable direct volume rendering of unstructured tetrahedral meshes, which we call DiffTetVR. The contributions of this work are as follows.

\begin{itemize}
    \item We derive equations for more accurate accumulation of color along view rays than simple alpha blending of locally sampled colors using ray marching (\cref{sec:fwd}).
    \item We introduce a fast rasterization-based renderer for tetrahedral meshes based on the concept of per-pixel linked lists~\cite{Yang2010} (\cref{sec:impl}).
    \item We deduce derivatives for the rendering process to enable backpropagation in the rendering pipeline regarding parameters such as vertex colors and positions (\cref{sec:bckwd,sec:bckwd-const}).
    \item We introduce a differentiable regularization term for avoiding degeneration of tetrahedral elements during the optimization process (\cref{sec:regularization}).
    \item We show how the tetrahedral mesh can be locally refined to enable a coarse-to-fine optimization approach (\cref{sec:tet-subdiv}).
    \item We demonstrate the reconstruction of pre-shaded volumes from images of scientific visualization volume data sets and synthetic surface renderings (\cref{sec:results}). We further evaluate strengths and weaknesses of the method (\cref{sec:discussion}).
\end{itemize}

\section{Related work}\label{sec:relwork}

\textbf{Differentiable rendering.}
As mentioned in the introduction, differentiable rendering has been used in previous works for reversing the rendering process for representations such as meshes~\cite{SoftRas,NvDiffRast}, point clouds~\cite{LinPointCloud} and 3D Gaussian splats~\cite{Kerbl3DGS}. A recent survey by Kato~\etal~\cite{DiffRenderSurvey} gives an overview of this wide field of works. The Mitsuba 2 rendering framework \cite{Mitsuba2} and Weiss and Westermann~\cite{DiffDVR} support differentiable volume rendering. The latter aims to solve the memory problems the former faces due to saving intermediate results for the backward pass.
While Weiss and Westermann only take into consideration volumetric absorption and emission, Nimier-David~\etal~\cite{nimierdavid2022unbiased} also take into account volumetric scattering using a differential ratio tracking \cite{novak14residual} approach. Similarly, Leonard~\etal~\cite{leonard2025lighttransportawarediffusionposterior} provide a differential volume renderer for single-view reconstruction of 3D volumes while supporting multiple scattering effects.
Shen~\etal~\cite{DMTet} introduced deep marching tetrahedra (DMTet), which builds upon the marching tetrahedra algorithm~\cite{MarchingTet} for surface reconstruction from images. Similarly, Yu~\etal~\cite{yu2024gaussianopacityfieldsefficient} also make use of the marching tetrahedra to extract a surface mesh from a 3DGS representation~\cite{Kerbl3DGS}.
Tetra-NeRF~\cite{TetraNeRF} generalizes the neural radiance field (NeRF) concept initially introduced by Mildenhall~\etal~\cite{NeRF} to use a dense tetrahedral grid constructed via Delaunay triangulation from an initial point cloud.
Gu~\etal~\cite{TetSplatting} introduce TeT-Splatting, which builds upon the work on 3DGS by Kerbl~\etal~\cite{Kerbl3DGS}. At every vertex of a tetrahedral mesh, they store the signed distance to the closest surface. Similar to our work, they interpolate these values using barycentric interpolation.
However, they do not aim to support differentiable volume rendering nor locally subdividing the mesh through the splitting of tetrahedral elements. In Appendix A.3 of their work, they describe the derivation of the gradient of the SDF values, but not the gradients wrt.\ the vertex positions. In \href{https://github.com/fudan-zvg/tet-splatting/blob/fd3d20dd8e151e1fca531a71fe631f574d7dbe7d/TetrahedronRender/src/tet_gradients.cu#L42}{their code}, however, they use a constant term for the position gradients that does not seem to take into account derivatives wrt.\ the barycentric coordinates. As this is not described in their work, it is unclear what the authors aim to achieve.

Govindarajan~\etal~\cite{RadiantFoam} introduce a method named \textit{Radiant Foam} that uses a polyhedral Voronoi tessellation of 3D space. While a Delaunay triangulation of a set of points is sensitive to edge flips, the authors describe that for the dual Voronoi diagram the ``shapes of the cells vary continuously with the positions of the points''. In our work, we do not re-tesselate the space after each optimization step, and aim to avoid edge flips by introducing a regularization term (cf.~\cref{sec:regularization}).

\textbf{Tet mesh rendering.}
Rendering volumetric tetrahedral meshes is a difficult task. The first question that arises is how to do optical accumulation of color within a single tetrahedral element. Williams and Max~\cite{TetIntegration} were the first to derive an exact formula for the accumulation of color along a ray in a tetrahedral element when using barycentric interpolation of color and opacity. We also provide a comparable formula in \cref{sec:fwd-exact}, but found the difference of two imaginary error functions to have a far too high numerical instability to be usable in real-world workflows on GPUs when using only 32-bit floating point precision. Thus, we derive a simplified formula with still higher accuracy than regular alpha blending in \cref{sec:fwd}.

Recent works have shown how ray tracing of tetrahedral meshes \cite{TetMeshRTShellTraversal,UnstructuredMeshSpaceSkip} and point location in tetrahedral meshes \cite{TetMeshPointLocationRTX} can be performed efficiently on modern hardware using specially crafted acceleration structures and hardware-accelerated ray tracing functionality. Historically used techniques often rely on rasterization-based techniques, especially the Painter's algorithm. The Painter's algorithm first sorts all polygons (or polyhedra) by their relative depth order before rasterizing them in this order. This depth order, however, can usually not be obtained by simply sorting by a linear depth value assigned to each element, but requires topological sorting~\cite{deBerg1993}. One example for a data structure that can be used for such a task is the binary space partitioning (BSP) tree~\cite{deBerg2008}. Chen~\etal~\cite{DepthPresortedTris} provide an approach for pre-sorting triangles and discuss a wide range of references on sorting directed acyclic graphs (DAGs), utilizing BSP trees and similar methods.
Williams~\cite{WilliamsMeshSortDAG} introduces a sorting technique for unstructured volumetric meshes with convex elements using directed acyclic graphs (DAGs), which they call the ``meshed polyhedra visibility ordering (MPVO)''. The concurrent work by Nelson~\etal~\cite{NelsonHanrahanCrawfisSort} describes a similar approach. A disadvantage of the MPVO technique is that a topological sort of the generated DAG is required every frame. Only few prior works have tackled the topic of parallel sorting of DAGs on GPUs, and usually assume certain constraints on the DAGs~\cite{GPUParTopSort}.
Shirley and Tuchman~\cite{ShirleyTuchmanDecomp} and Kraus~\etal~\cite{ProjectingTetrahedra} describe a technique called \textit{projected tetrahedra (PT)}, where they project tetrahedra to a set of triangular faces that can then be rasterized. Kraus~\etal~\cite{ProjectingTetrahedra} improve the technique by ensuring perspective correct interpolation without rendering artifacts. The PT approach suffers from the same problem as directly rasterizing the tetrahedra, which is that topological sorting is required. Stein~\etal~\cite{SteinExp} extend the work by Shirley and Tuchmann to take into account that non-linear terms such as $e^{-\alpha t}$ (for an opacity $\alpha$ and a segment length $t$) introduced by the Beer-Lambert law should not be interpolated linearly. This is also utilized in the implementation of PT in the VTK toolkit~\cite{vtkBook}, where the segment length is interpolated linearly instead of interpolating the non-linear terms.


\textbf{Order-independent transparency (OIT).}
As discussed in the last paragraphs, the difficulty to implement parallel topological sorting on GPUs makes the Painter's algorithm hard to use for efficient rendering. While the z-buffer has made the Painter's algorithm obsolete for rendering opaque meshes on modern graphics hardware, it can only be used to resolve the closest visible surface. The A-buffer~\cite{ABuffer} aims to overcome some of the restrictions of the z-buffer, but cannot be implemented as-is on modern graphics hardware. Yang~\etal~\cite{Yang2010} build on the idea of the A-buffer and describe an algorithm for storing per-pixel linked lists to gather generated fragments in a first rendering pass before sorting them per-pixel in a second rendering pass. A disadvantage of this kind of technique is the potentially unbounded memory requirement. Depth peeling~\cite{Everitt2001,Bavoil2008}, on the other hand, trades the unbounded memory requirement for needing as many rendering passes as there are elements on top of each other (called the \textit{depth complexity} of a scene). Approximate techniques such as multi-layer alpha blending (MLAB)~\cite{Salvi2014}, moment-based order-independent transparency (MBOIT)~\cite{Muenstermann2018} or stochastic transparency~\cite{StochasticTransparency} trade potential rendering inaccuracies for bounded memory requirements and a bounded number of render passes.

\newpage
\section{Forward rendering}\label{sec:fwd}

Given a tetrahedral mesh with vertex colors, opacities and positions, we want to compute for each pixel of a virtual camera the final color of the view ray passing through the mesh. Like Weiss and Westermann~\cite{DiffDVR}, we only take into consideration absorption and emission effects in direct volume rendering and ignore volumetric scattering. For each tetrahedron the ray intersects with, the color is accumulated using \cref{eq:accum}. We assume a function $\lerp(val_0, val_1, x) = (1 - x) \cdot val_0 + x \cdot val_1$ is given that linearly interpolates between two values. $c_0, \alpha_0$ and $c_1, \alpha_1$ are the color and opacity at the entry and exit face obtained using barycentric interpolation.

\begin{equation}
\label{eq:accum}
\begin{aligned}
c_{accum}(t, c_0, c_1, \alpha_0, \alpha_1) &= \int_0^t \text{lerp}(c_0, c_1, x) \text{lerp}(\alpha_0, \alpha_1, x) e^{-\int_0^x \text{lerp}(\alpha_0, \alpha_1, u) \,du} \,dx \\
&= \int_0^t \left(\left(1-x\right)c_0 + x c_1\right)\left(\left(1-x\right)\alpha_0 + x \alpha_1\right) e^{-\int_0^x \left(1-u\right)\alpha_0 + u \alpha_1 \,du} \,dx \\
&= \int_0^t \left( r_0 + r_1 x + r_2 x^2 \right) e^{q_1 x + q_2 x^2} \,dx
\end{aligned}
\end{equation}

$r_0$, $r_1$, $r_2$, $q_1$ and $q_2$ are five coefficients obtained from $c_0, \alpha_0$ and $c_1, \alpha_1$. A challenge in solving \cref{eq:accum} is the term $q_2 x^2$ in the exponent, as $\int e^{q_2 x^2} \,dx$ is considered a nonelementary integral. In \cref{sec:fwd-exact}, we provide a full derivation of \cref{eq:accum} based on the error function $\erf(z)$ and the imaginary error function $\ierf(z)$, but we found it generally not suitable for actual implementation on a GPU. Firstly, while the Bürmann series can be used as a numerically stable approximation of the error function~\cite{schopf2014burmann}, we were not able to implement a fast and numerically stable approximation for the imaginary error function. The latter might be due to $\lim_{z \to \infty} \erf(z) = 1$, but $\lim_{z \to \infty} \ierf(z) = \infty$ and additional cancellation effects due to the subtraction of large floating point values. With realistic values for the opacities, the evaluations of $\text{erfi}$ may assume values of a magnitude of $10^{20}$. Finally, the partial derivatives for the colors and opacities become very convoluted. Consequently, we have decided to simplify \cref{eq:accum} by assuming that $\alpha_0$ equals $\alpha_1$. In this case, both $q_2 x^2$ and $r_2 x^2$ vanish and the integral can be expressed using elementary functions (see \cref{alg:linear-accum}). In order to be able to use this approximation, we separate the whole interval from \cref{eq:accum} into multiple sub-intervals and assume that the linearly interpolated opacity at the center is constant over the sub-interval.

In \cref{alg:tet-col}, it is shown how the computation of the accumulated color within a tetrahedral cell is performed. It relies on \cref{alg:bary} for barycentric interpolation and \cref{alg:linear-accum} for accumulation along a sub-interval with constant opacity.

\begin{algorithm}
\caption{Tetrahedron color accumulation}
\label{alg:tet-col}
\SetKwInOut{Input}{Input}
\SetKwInOut{Parameters}{Parameters}
\Input{Face indices $f_{0}$ and $f_{1}$ of hit tet and corresponding depths $d_0$, $d_1$, previously accumulated color and opacity $C_{ray}$, $\alpha_{ray}$, number of subdivision steps $N_{sub}$}
\Parameters{Tet vertex positions $P_{T}$, vertex colors $C_{T}$}
$p_{0}, c_{0}$ = \textbf{Bary}($d_0$, $f_0$)\;
$p_{1}, c_{1}$ = \textbf{Bary}($d_1$, $f_1$)\;
$t_{total}$ = $\lVert p_{1} - p_{0} \rVert_2$\;
$t$ = $t_{total} / N_{sub}$\;
$c_{ray,0}$ = $c_{ray}$\;
$\alpha_{ray,0}$ = $\alpha_{ray}$\;
\For{$i = 0$ \KwTo $N_{sub} - 1$}
{
  $c_{i0}$ = \textbf{lerp}($c_0$, $c_1$, $i/N_{sub}$)\;
  $c_{i1}$ = \textbf{lerp}($c_0$, $c_1$, $(i+1)/N_{sub}$)\;
  $\alpha_{i}$ = \textbf{lerp}($\alpha_0$, $\alpha_1$, $(i+0.5)/N_{sub}$)\;
  $c_{acc,i}$, $\alpha_{acc,i}$ = \textbf{Accum}($t$, $c_{i0}$, $c_{i1}$, $\alpha_{i}$)\;
  $c_{ray,i+1} = c_{ray,i} + (1 - \alpha_{ray,i}) c_{acc,i}$\;
  $\alpha_{ray,i+1} = \alpha_{ray,i} + (1 - \alpha_{ray,i}) \alpha_{acc,i}$\;
}
\textbf{return} $c_{ray,N_{sub}}$, $\alpha_{ray,N_{sub}}$\; 
\end{algorithm}

\begin{algorithm}
\caption{Barycentric interpolation}
\label{alg:bary}
\SetKwInOut{Input}{Input}
\SetKwInOut{Parameters}{Parameters}
\textbf{Bary} $(d, f)$\;
\Input{Hit depth $d$, tet face index $f$, screen position gl\_FragCoord.xy, global tet face vertex table $F: \{0, \dots, N_f - 1\} \to \{0, \dots, N_v - 1\}^3$}
\Parameters{Tet vertex positions $P_{T}$, vertex colors $C_{T}$}
\tcc{NDC: Normalized device coordinates}
$p_{NDC}$ = $(2 * \text{gl\_FragCoord.xy} / \text{viewportSize} - 1, d, 0)$\;
$p_H$ = $\text{invViewProjMat} \cdot p_{NDC}$\;
$p$ = $p_{H,xyz} / p_{H,w}$\;
$i_{0}$, $i_{1}$, $i_{2}$ = $F(f)$\;
$p_{0}$ = $P_T(i_{0})$, $p_{1}$ = $P_T(i_{1})$, $p_{2}$ = $P_T(i_{2})$\;
$c_{0}$ = $C_T(i_{0})$, $c_{1}$ = $C_T(i_{1})$, $c_{2}$ = $C_T(i_{2})$\;
$area$ = $\lVert(p_{2} - p_{0}) \times (p_{2} - p_{1})\rVert_2$\;
$u$ = $\lVert(p_{2} - p_{1}) \times (p_{2} - p_{1})\rVert_2 / area$\;
$v$ = $\lVert(p_{2} - p_{0}) \times (p_{2} - p_{1})\rVert_2 / area$\;
$c = c_{0} \cdot u + c_{1} \cdot v + c_{2} \cdot (1 - u - v)$\;
\tcc{for gradient: $p = p_{0} \cdot u + p_{1} \cdot v + p_{2} \cdot (1 - u - v)$}
return $p$, $c$\;
\end{algorithm}

\begin{algorithm}
\caption{Linear color accumulation}
\label{alg:linear-accum}
\SetKwInOut{Input}{Input}
\SetKwInOut{Parameters}{Parameters}
\textbf{Accum} $(t, c_0, c_1, \alpha)$\;
\Input{Optical depth $t$, entry color $c_0$, exit color $c_1$, constant opacity $\alpha$}
$A$ = $e^{-\alpha t}$\;
$c_{acc}$ = $\left(1 - A\right) \cdot c0 + \left(\left(t + \frac{1}{\alpha}\right) \cdot A - \frac{1}{\alpha}\right) \cdot \left(c0 - c1\right)$\;
$\alpha_{acc}$ = $1 - A$\;
return $c_{acc}$, $\alpha_{acc}$\;
\end{algorithm}

\newpage
\section{Implementation}\label{sec:impl}

As outlined in the last section, we need to find the intersections of all tetrahedra with the view rays of a virtual camera for all rendered pixels.
While it is possible to use raycasting for determining the intersections of the view rays with the tetrahedral cells, a disadvantage is that an acceleration structure needs to be built for fast rendering, which becomes invalid in each iteration of the optimizer when the vertex positions of the mesh are updated.
Another alternative would be using the Painter's algorithm described in \cref{sec:relwork}, but as outlined in that section, to our knowledge no efficient parallel DAG sorter implementation exists for GPUs.
Thus, we use per-pixel linked lists~\cite{Yang2010} to first rasterize the faces of the tetrahedra and then sort the intersections of the view rays in front-to-back order. We make use of the GPU-friendly heap-sort implementation provided by Kern~\etal~\cite{TransparentLines}. In the backward pass, the intersections are sorted in reverse order to enable backpropagation using the inversion trick by Weiß and Westermann~\cite{DiffDVR}. Our implementation is provided in a software framework using the Vulkan graphics API \cite{VulkanSpec} at \url{https://github.com/chrismile/DiffTetVR}. Vulkan has the advantage over pure compute-based APIs like NVIDIA CUDA that it supports hardware rasterization. Thus, it is not necessary to develop a (likely slower) tile-based software rasterizer as was done, for example, by Kerbl~\etal~\cite{Kerbl3DGS}. We use Vulkan-CUDA API interop to share buffers and semaphores between CUDA and Vulkan when accessing data via PyTorch bindings.

A final question is how to efficiently store unstructured meshes in memory.
Kremer~\etal~\cite{OpenVolumeMeshPaper} provide a software library called OpenVolumeMesh that enables the efficient storage and manipulation of unstructured meshes, which we make use of in our work.
In \cref{sec:tet-subdiv}, we show how we use the prism tesselation technique described by Erleben~\etal~\cite{AdaptiveThinShellTetMesh} in order to subdivide tetrahedra during the optimization process. Given a triangular mesh as a boundary representation, they aim to construct a so-called thin shell tetrahedral mesh. For this, the triangles are extruded to form prisms, which are then tessellated into tetrahedra. We utilize OpenVolumeMesh for efficiently adding vertices, edges and tetrahedra to the underlying data structure.

\newpage
\section{Backward rendering}\label{sec:bckwd}

We assume that a target image and camera intrinsics and extrinsics are given. During backward rendering, the loss between the target image and the image obtained using the forward renderer is computed. Our software framework currently supports an $l_1$ or $l_2$ loss, but support for arbitrary losses can be implemented. By differentiating the loss, gradients $c_{adj}$, $\alpha_{adj}$ for all input pixel colors and opacities are obtained. In order to backpropagate the gradients through the renderer, the generated fragments are no longer sorted front-to-back like in the forward pass, but back-to-front. Hits with the tetrahedra and the discretized accumulations are computed in reverse order. Our goal is to obtain gradients for the tetrahedron vertex positions, colors and opacities. Consequently, gradients need to be backpropagated using the chain rule for any variable depending on one of those attributes.


\begin{equation}
c_{acc}(t, c_0, c_1, \alpha) = \left( 1 - e^{-\alpha t} \right) c_0 + \left( \left( t + \frac{1}{\alpha} \right) e^{-\alpha t} - \frac{1}{\alpha}\right) \left( c_0 - c_1 \right)
\end{equation}
\begin{equation}
a_{acc}(t, \alpha) = 1 - e^{-\alpha t}
\end{equation}
\begin{equation}
\frac{\partial c_{acc}(t, c_0, c_1, \alpha)}{\partial c_0} = \left( 1 - e^{-\alpha t} \right) + \left( \left( t + \frac{1}{\alpha} \right) e^{-\alpha t} - \frac{1}{\alpha}\right)
\end{equation}
\begin{equation}
\frac{\partial c_{acc}(t, c_0, c_1, \alpha)}{\partial c_1} = -\left( \left( t + \frac{1}{\alpha} \right) e^{-\alpha t} - \frac{1}{\alpha}\right)
\end{equation}
\begin{equation}
\frac{\partial c_{acc}(t, c_0, c_1, \alpha)}{\partial \alpha} = t e^{-\alpha t} c_0 + \left( \frac{1}{\alpha^2} - \left( t^2 + \frac{1}{\alpha^2} + \frac{t}{a} \right) e^{-\alpha t} \right) \left( c_0 - c_1 \right)
\end{equation}
\begin{equation}
\frac{\partial c_{acc}(t, c_0, c_1, \alpha)}{\partial t} = \left( \alpha c_0 + \left( \alpha c_1 - \alpha c_0 \right) t \right) e^{-\alpha t}
\end{equation}
\begin{equation}
\frac{\partial a_{acc}(t, \alpha)}{\partial \alpha} = t e^{-\alpha t}
\end{equation}
\begin{equation}
\frac{\partial a_{acc}(t, \alpha)}{\partial t} = \alpha e^{-\alpha t}
\end{equation}

Furthermore, for $\alpha = 0$ with application of L'Hôpital's rule:

\begin{equation}
\lim_{\alpha \to 0} \frac{\partial c_{acc}(t, c_0, c_1, \alpha)}{\partial \alpha} = t c_0 + \frac{1}{2} t^2 \left( c_1 - c_0 \right)
\end{equation}
\begin{equation}
\lim_{\alpha \to 0} \frac{\partial c_{acc}(t, c_0, c_1, \alpha)}{\partial c_0} = \lim_{\alpha \to 0} \frac{\partial c_{acc}(t, c_0, c_1, \alpha)}{\partial c_1} = 0
\end{equation}

We use the inversion trick by Weiß and Westermann~\cite{DiffDVR}.

\begin{equation}
  c_{ray,i+1} = c_{ray,i} + (1 - \alpha_{ray,i}) c_{acc,i}
\end{equation}
\begin{equation}
  \alpha_{ray,i+1} = \alpha_{ray,i} + (1 - \alpha_{ray,i}) \alpha_{acc,i}
\end{equation}
\begin{equation}
  \Rightarrow \alpha_{ray,i} = \frac{\alpha_{acc,i} - \alpha_{ray,i+1}}{\alpha_{acc,i} - 1} 
\end{equation}
\begin{equation}
  \Rightarrow c_{ray,i} = c_{ray,i+1} - (1 - \alpha_{ray,i}) c_{acc,i}
\end{equation}

The analytic derivatives of the barycentric coordinates with respect to the vertex positions have been automatically deduced using the SymPy library~\cite{SympyPaper} and can be found in the accompanying source code in the file \href{https://github.com/chrismile/DiffTetVR/blob/main/Data/Shaders/Common/BackwardCommon.glsl}{\texttt{Data/Shaders/Common/BackwardCommon.glsl}} together with the derivatives of the optical depth $t$ of the view ray within a tetrahedron.

\newpage
\section{Regularization}\label{sec:regularization}

\begin{figure*}[t]
\centering
\includegraphics[height=3.1cm]{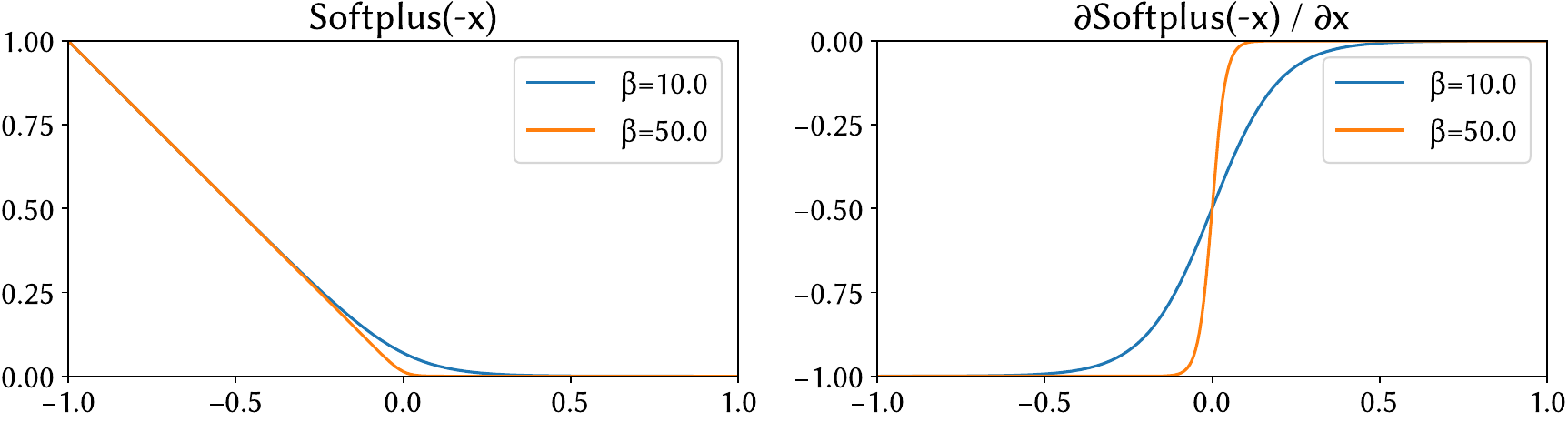}
\caption{Used function for mapping the tetrahedral element quality to the regularization penalty and its derivative.}
\label{fig:softplus}
\end{figure*}

\begin{figure*}[t]
\centering
\captionsetup[subfigure]{labelformat=empty}
\subfloat[\label{subfig:regularizer-0}0 steps]{\includegraphics[width=0.15\linewidth]{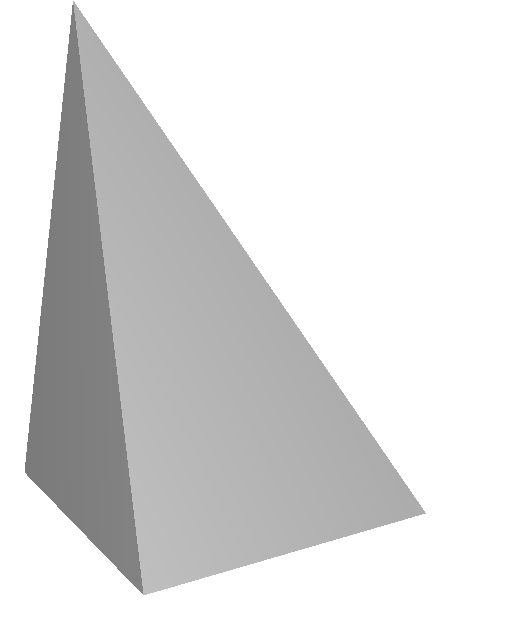}}\hspace{0.02\linewidth}
\subfloat[\label{subfig:regularizer-100}100 steps]{\includegraphics[width=0.15\linewidth]{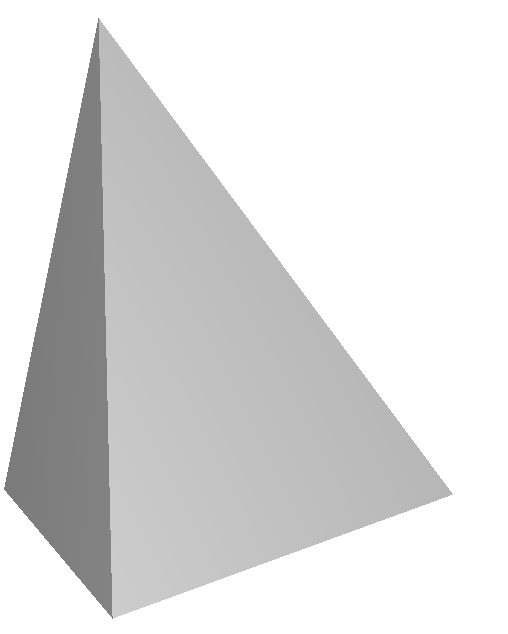}}\hspace{0.02\linewidth}
\subfloat[\label{subfig:regularizer-200}200 steps]{\includegraphics[width=0.15\linewidth]{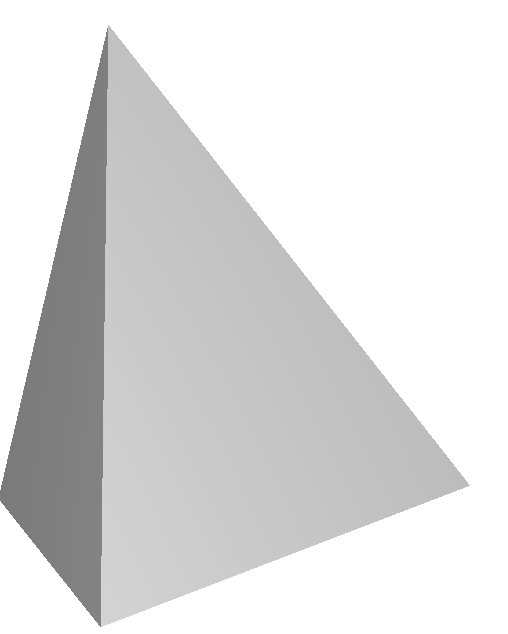}}\hspace{0.02\linewidth}
\subfloat[\label{subfig:regularizer-600}600 steps]{\includegraphics[width=0.15\linewidth]{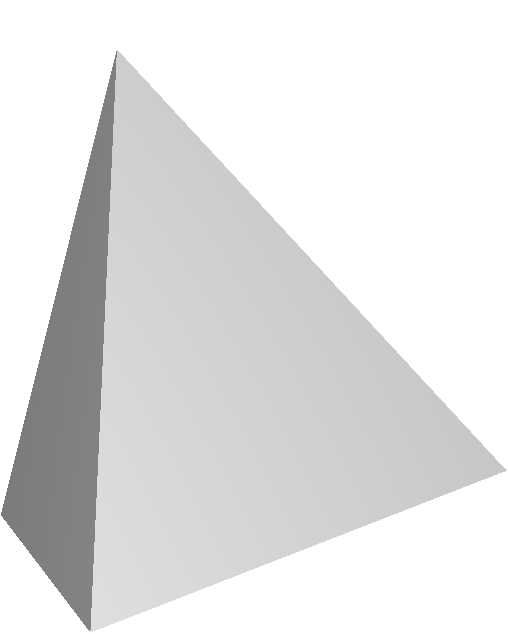}}
\caption{Regularizer ($\beta=10$) applied to a non-regular tetrahedron. Setting a higher value for $\beta$ penalizes only slightly deformed tetrahedra less (cf.~\cref{fig:softplus}).}
\label{fig:regularizer}
\end{figure*}

When rendering tetrahedral meshes, our face-based rendering technique requires that self-intersections may not happen and that tetrahedral elements are not inverted, i.e., their volume may not become negative. Also, tetrahedral elements should not degenerate, i.e., their volume should not approach zero. This can happen during the optimization process due to the $C0$-continuity of the color field and when using too high learning rates.

Shewchuk~\cite{TetQualityShort,TetQualityFull} discusses different choices of quality measures for triangles and tetrahedra. These measures are derived in different ways such that they fulfill certain properties with respect to interpolation errors or stiffness matrix conditioning. For tetrahedra, they are based on individual sub-measures such as the signed volume, the face areas and the edge lengths. Shewchuck says that ``smooth measures simplify optimization-based smoothing, but they are based on weaker bounds, so they are less accurate indicators than the nonsmooth measures''~\cite{TetQualityFull}. In our application, we are not directly interested in tet regularization due to interpolation properties themselves, but to add a regularization term to the optimization process that helps to avoid tetrahedral element degeneration. For us, the smoothness property is thus of greater importance. For our regularizer, we have chosen the regularization term $ \text{Softplus}(-Q)$ using the smooth measure $Q$ in \cref{eq:quality}, which Shewchuk~\cite{TetQualityFull} attributes to Parthasarathy~\etal~\cite{Parthasarathy}. As shown in \cref{fig:softplus}, a regularizer using the softplus term penalizes inverted and degenerate tetrahedra with negative or close to zero volume. \cref{fig:regularizer} shows the effect of the regularizer (without any other image-based loss) on a tetrahedron after multiple iterations. The tetrahedron slowly approaches a regular tetrahedron (i.e., the four faces become equilateral triangles). Depending on the softplus parameter $\beta$ and the weight $\lambda$ assigned to the regularizer in the complete loss term, only strongly degenerate tetrahedra can be penalized and the penalization factor can be reduced.

\begin{equation}
\label{eq:quality}
Q = 6 \sqrt{2} \frac{V}{l_{rms}^3}
\end{equation}

\begin{equation}
V = \langle (p_3 - p_0), (p_1 - p_0) \times (p_2 - p_0) \rangle
\end{equation}

\begin{equation}
l_{rms} = \sqrt{\frac{1}{6}\sum_{i=0}^{5} l_i^2}
\end{equation}

\begin{equation}
\text{Softplus}(x) = \frac{1}{\beta} \log(1 + \exp(\beta x))
\end{equation}

\begin{equation}
\frac{\partial \text{Softplus}(x)}{\partial x} = \frac{1}{1 + \exp(-\beta x)}
\end{equation}

The analytic derivatives of the quality metric $Q$ with respect to the tet vertex positions $p_{\{0,1,2,3\}}$ has been automatically deduced using the SymPy library~\cite{SympyPaper} and can be found in the accompanying source code in the file \href{https://github.com/chrismile/DiffTetVR/blob/main/Data/Shaders/Optimizer/TetRegularizer.glsl}{\texttt{Data/Shaders/Optimizer/TetRegularizer.glsl}}. The complete loss term is given below, where $T$ is the set of all tetrahedra.

\begin{equation}
\text{loss} = \text{loss}_{\text{img}} + \lambda \sum_{t \in T} \text{Softplus}(Q(t))
\end{equation}

\newpage
\section{Tet subdivision}\label{sec:tet-subdiv}

\begin{figure*}[t]
\centering
\includegraphics[width=0.75\linewidth]{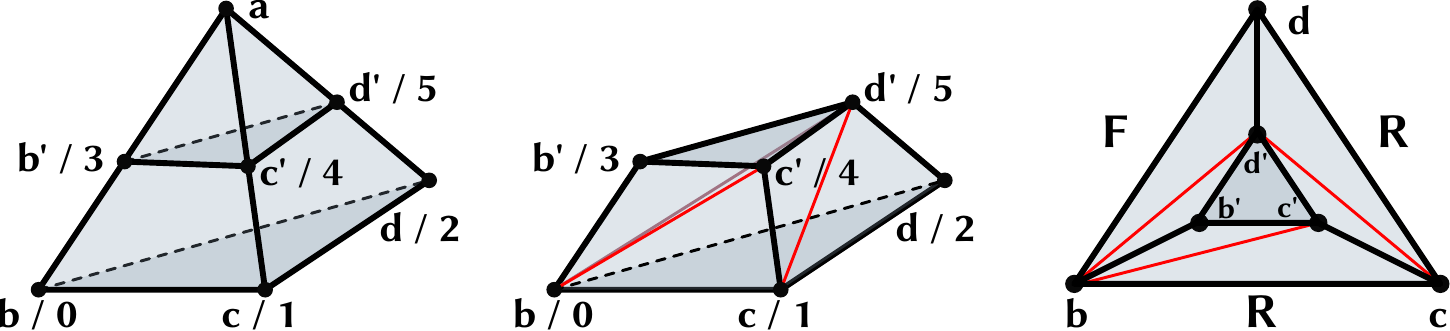}
\caption{Left: All edges incident to vertex $a$ are split. Middle: The incident tetrahedra are split by first removing the newly formed tetrahedron incident to $a$ and then splitting the remaining truncated pyramid in the form of a triangular prism formed by $b$, $c$, $d$, $b'$, $c'$ and $d'$. Right: Top-down view indicating how the quad faces of a truncated pyramid prism are split using rising (\textbf{R}) and falling (\textbf{F}) edges.}
\label{fig:prism-split}
\end{figure*}

\begin{figure*}[t]
\centering
\captionsetup[subfigure]{labelformat=empty}
\subfloat[\label{subfig:quad-split-a}Compatible splits]{\includegraphics[width=0.2\linewidth]{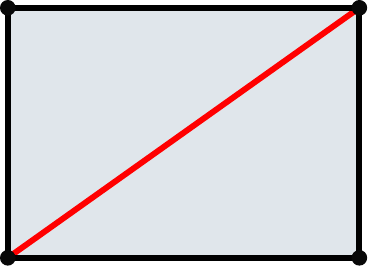}}\hspace{0.02\linewidth}
\subfloat[\label{subfig:quad-split-b}Incompatible splits]{\includegraphics[width=0.2\linewidth]{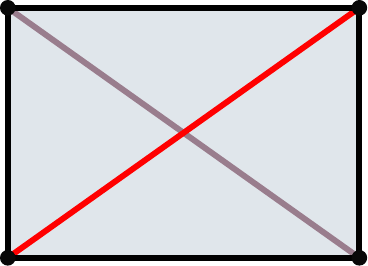}}\hspace{0.02\linewidth}
\caption{Left: The split of a quad face incident to two adjacent truncated pyramid prisms are compatible. Right: The splits are in opposite directions and thus incompatible.}
\label{fig:quad-split}
\end{figure*}

When optimizing the properties of the vertices of a tetrahedral mesh, a coarse-to-fine approach can be chosen for better convergence. In this case, the optimization process is started with a relatively coarse mesh. When accumulating gradients over multiple images and corresponding camera poses, some vertices may have larger gradients than others. This may indicate an insufficient local resolution of the mesh that is unable to represent certain patterns of high-frequency scales. In this case, we may decide to refine and subdivide the mesh only in the vicinity of these vertices. Similarly to AbsGS~\cite{AbsGS}, we also found that absolute gradients result in better results for splitting.

Consequently, we propose the following approach. After the gradients have been accumulated, we select a set of $\hat{N}$ vertices with the largest gradient magnitudes. Then, we iterate over all of these vertices and split all incident edges by inserting a new vertex at, e.g., the middle of each edge. However, this also requires splitting the incident tetrahedra using the newly inserted vertices.
As indicated in \cref{fig:prism-split}, this splitting operation results in a new tetrahedron incident with the splitting vertex $a$ and a truncated pyramid in the form of a triangular prism.

\begin{figure*}[t]
\centering
\includegraphics[width=0.9\linewidth]{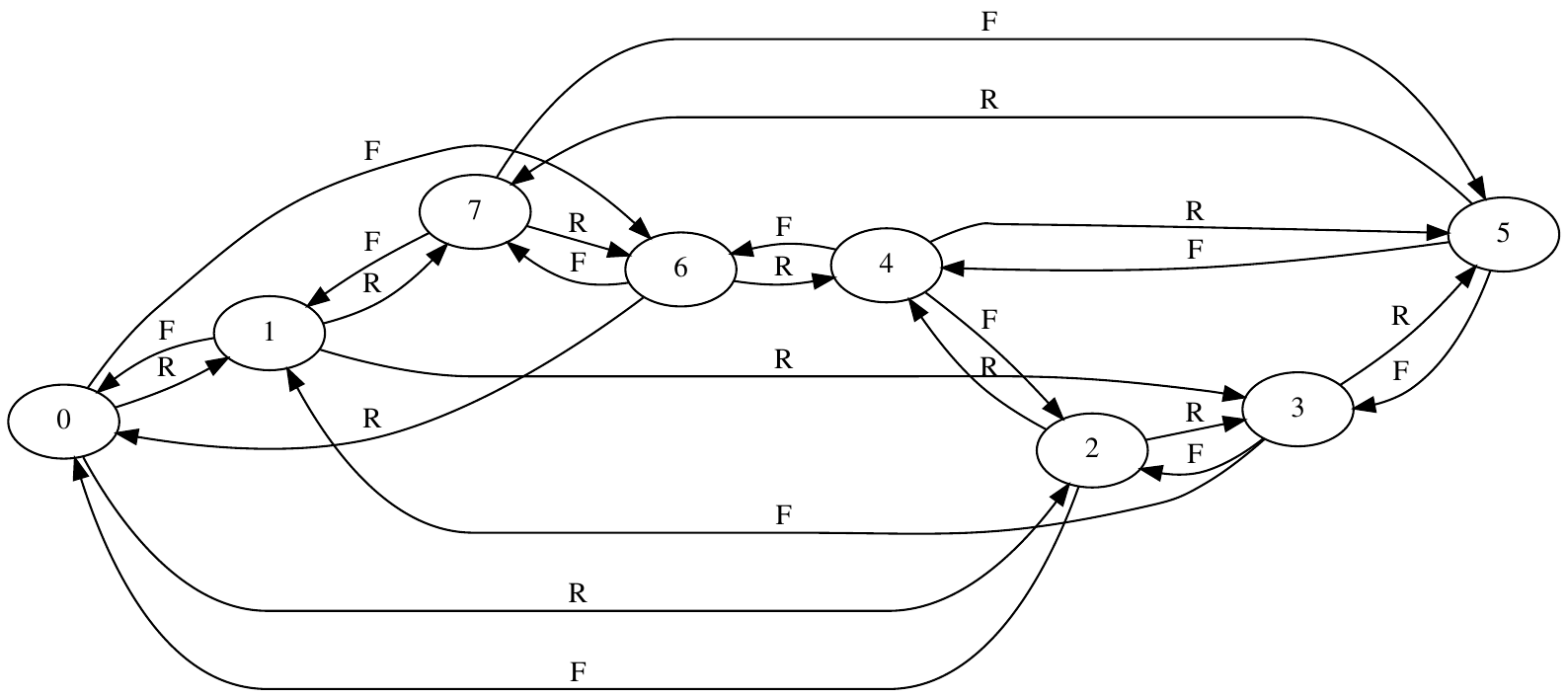}
\caption{Dual graph of a solved CSP problem. The nodes represent prisms and the edges the splits of shared incident quad faces.}
\label{fig:csp-example}
\end{figure*}

Erleben~\etal~\cite{AdaptiveThinShellTetMesh} describe an algorithm for splitting a triangular prism mesh into a tetrahedral mesh in section 4 of their work. Each triangular prism can be split into exactly three tetrahedra. However, care needs to be taken so that the face splits are compatible with the splits of adjacent tetrahedra. As indicated in \cref{fig:quad-split}~right, when adjacent prisms sharing a quad face decide for incompatible splitting directions of the quad face, the triangle faces of the subdivided tetrahedra cannot be shared, which leads to an infinitesimally thin hole in the underlying tet mesh. Erleben~\etal categorize split edges as falling (\textbf{F}) or rising (\textbf{R}) depending on ``whether the tesselation edge is falling or rising as we travel along the extruded prism face in counter clock wise manner'' \cite{AdaptiveThinShellTetMesh}. Two splits of a shared quad face are compatible if one of them is a rising edge split and the other one is a falling edge split. This leads to a constraint satisfaction problem (CSP) with the following constraints:
\begin{itemize}
    \item The splits for one prism may not be \textbf{FFF} or \textbf{RRR}.
    \item Two adjacent prisms must split a shared quad face using exactly one \textbf{F} and one \textbf{R} split.
\end{itemize}
While this CSP could be solved with general constraint satisfaction libraries, Erleben~\etal give a specialized algorithm with pseudo-code for solving this problem. Unfortunately, their pseudo-code leaves open many implementation details. In our code, we provide an implementation of their approach in the file \href{https://github.com/chrismile/DiffTetVR/blob/main/src/Tet/CSP/FlipSolver.cpp}{\texttt{FlipSolver.cpp}}. In \cref{tab:split-lookup}, we provide a tesselation look-up table for the vertex indices as assigned in \cref{fig:prism-split} and specific tesselation patterns. An example for a solved CSP problem dual graph can be found in \cref{fig:csp-example}.

For the initial tetrahedral mesh, the user can either specify to use a regular grid aligned with the bounding box of the scene to be optimized, or to tetrahedralize the sparse point cloud generated by the structure from motion (SfM) algorithm provided by, e.g., COLMAP~\cite{schoenberger2016sfm,schoenberger2016mvs}. For this, a Delaunay triangulation is used. The regular grid tetrahedral mesh can either be generated by converting the hexahedral mesh cells separately to tetrahedral cells, or by using the meshing libraries fTetWild~\cite{fTetWild} or TetGen~\cite{TetGen}.

\begin{table*}[t]
\centering
\begin{tabularx}{\textwidth}{ |c| *{6}{>{\centering\arraybackslash}X|} }
\cline{2-7}
   \multicolumn{1}{c|}{} 
 & \multicolumn{6}{c|}{Prism splits}\\
\hline
 Tet index & FRR & RFR & FFR & RRF & FRF & RFF\\
\hline
0 & \{0, 1, 3, 2\} & \{0, 1, 4, 2\} & \{0, 1, 3, 2\} & \{0, 1, 5, 2\} & \{0, 1, 5, 2\} & \{0, 1, 4, 2\} \\
1 & \{3, 4, 5, 1\} & \{3, 4, 5, 2\} & \{3, 4, 5, 2\} & \{3, 4, 5, 0\} & \{3, 4, 5, 1\} & \{3, 4, 5, 0\} \\
2 & \{1, 2, 5, 3\} & \{0, 2, 4, 3\} & \{1, 2, 4, 3\} & \{0, 1, 4, 5\} & \{0, 1, 3, 5\} & \{0, 2, 4, 5\} \\
\hline
\end{tabularx}
\caption{Look-up table for splitting a triangular prism using different splitting edge patterns. The indices correspond to prism vertices as specified in \cref{fig:prism-split}.}
\label{tab:split-lookup}
\end{table*}

\newpage


\begin{figure*}[t]
\centering
\includegraphics[width=0.324\linewidth]{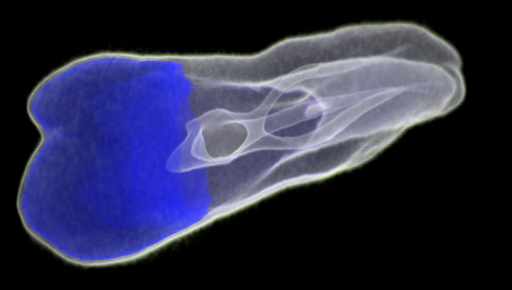}\hspace{0.05cm}
\includegraphics[width=0.324\linewidth]{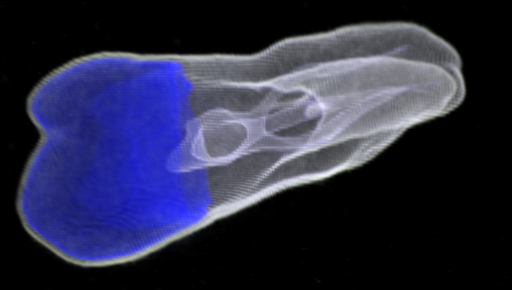}\hspace{0.05cm}
\includegraphics[width=0.324\linewidth]{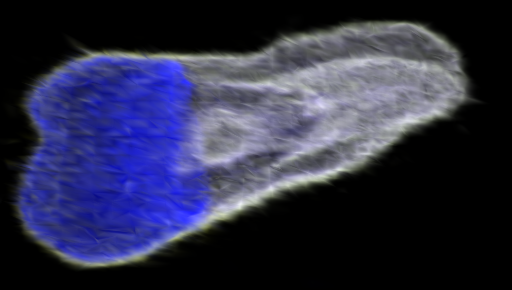}
\caption{Left: Ground truth rendering of \textit{Tooth} data set. Middle: Color-only tet mesh reconstruction with a regular grid of $177 \times 100 \times 100$ converted into a tetrahedral mesh ($lr_c=0.08$, PSNR $33.2$). Right: Joint color and position optimization ($\text{lr}_c=0.08$, $\text{lr}_p=10^{-6}$, $\lambda=10$, $\beta=100$, PSNR $26.3$).}
\label{fig:tooth-img}
\end{figure*}

\begin{figure*}[t]
\centering
\includegraphics[width=0.45\linewidth]{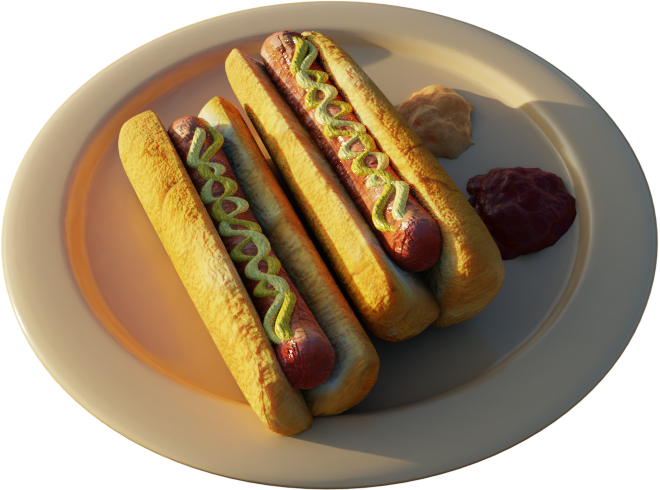}\hspace{0.5cm}
\includegraphics[width=0.45\linewidth]{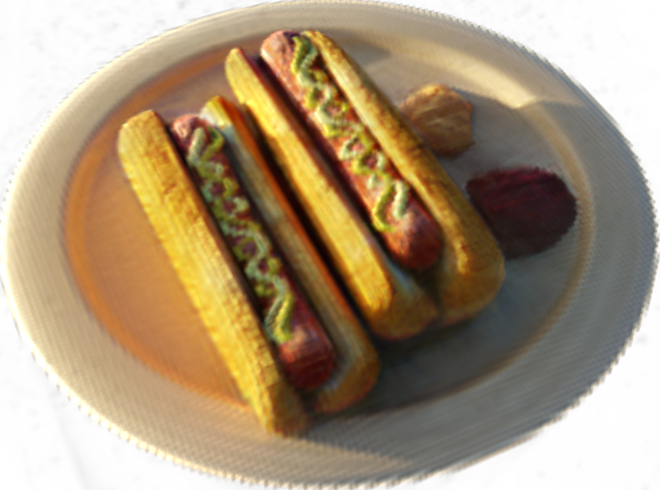}
\caption{Training for synthetic \textit{Hotdog} data set. Left: Ground truth test set rendering~\cite{NeRF}. Right: Tetrahedral mesh reconstruction using a regular grid of size $128^3$ converted into a tetrahedral mesh ($\text{lr}_c=0.08$, PSNR 25.2). Without adaptive subdivision, most tetrahedra are fully transparent.}
\label{fig:hotdog-img}
\end{figure*}

\begin{figure*}[t]
\centering
\includegraphics[width=0.48\linewidth]{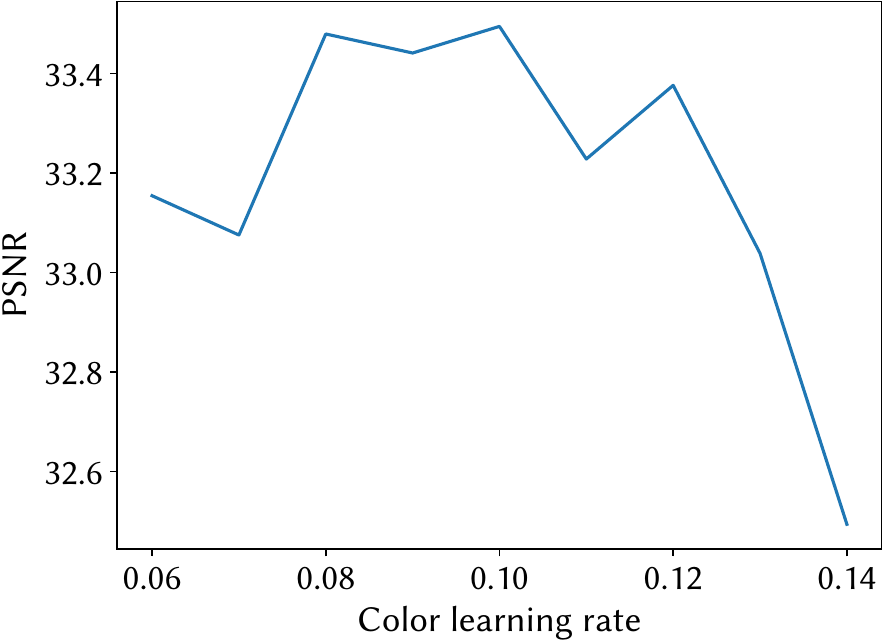}\hspace{0.5cm}
\includegraphics[width=0.48\linewidth]{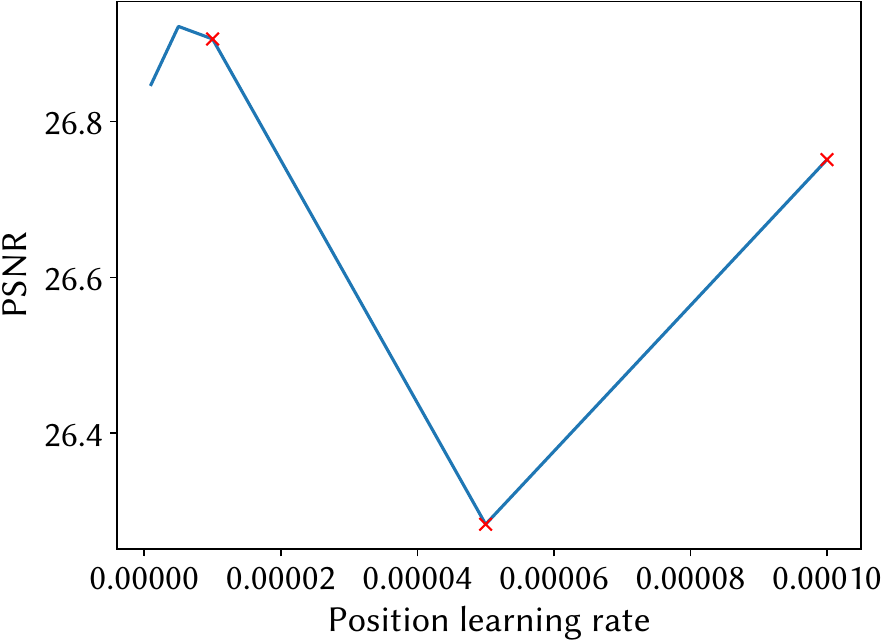}\\
\includegraphics[width=0.48\linewidth]{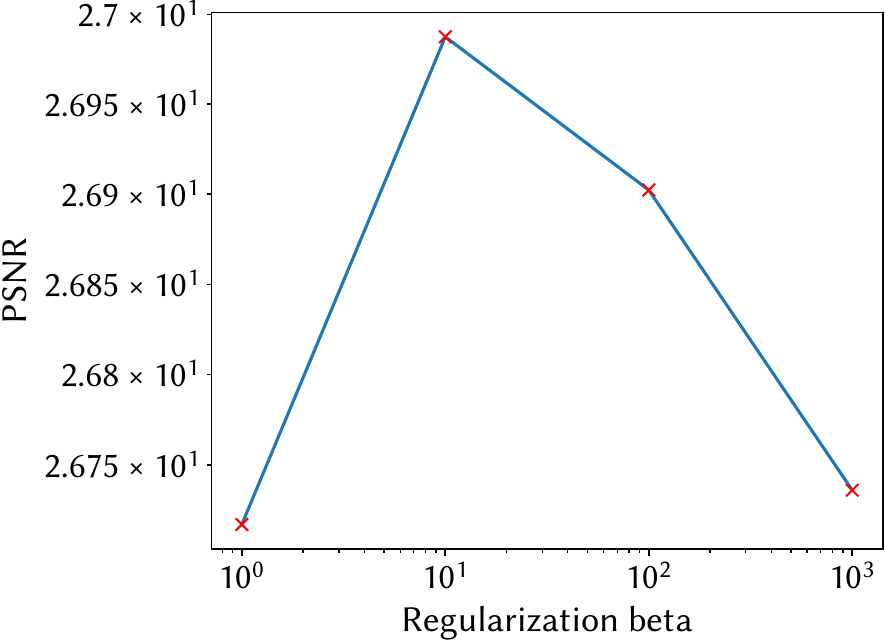}\hspace{0.5cm}
\includegraphics[width=0.48\linewidth]{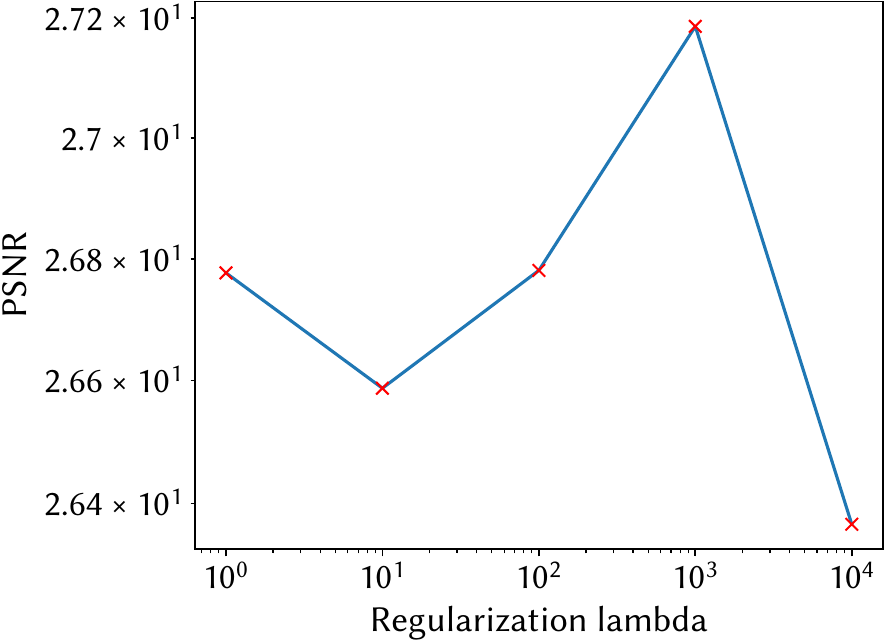}
\caption{Hyperparameter analysis for reconstructing a preshaded volume for the \textit{Tooth} data set. A red "X" indicates that the final tetrahedral mesh contained degenerate elements with volume $\leq 0$. Top left: The color learning rate peaked at about 33.5 PSNR for $[0.08, 0.1]$. Top right: Training with position learning rate $\neq 0$ lead to stable results only for the lowest learning rates (albeit with lower PSNR than color-only optimization). Bottom: Fine-tuning regularization parameters as outlined in \cref{sec:regularization} did not lead to a significant improvement. Regularization not able to stabilize results at $\text{lr}_p=10^{-5}$ to result in tet meshes with non-degenerate elements. Unspecified hyperparameters can be found \href{https://github.com/chrismile/DiffTetVR/blob/0643199f745c98895ae973b40a99ee7462d945b3/pytests/run.py\#L114}{online}.}
\label{fig:tooth-dia}
\end{figure*}

\section{Results}\label{sec:results}

Weiss and Westermann~\cite{DiffDVR} have demonstrated how DiffDVR can be used to reconstruct transfer functions, voxel densities and pre-shaded volumes from pre-rendered images. In this section, we evaluate the ability of DiffTetVR to not only accurately reconstruct pre-shaded vertex colors from images rendered with an emission-absorption volume rendering model, but to also optimize the number of overall tetrahedral elements and their placement. Unlike DiffDVR, DiffTetVR can use a coarse-to-fine optimization approach where more elements are placed in regions with higher reconstruction error. For this, the technique described in \cref{sec:tet-subdiv} is used. All tests were performed on an NVIDIA RTX 3090 GPU and a 12-core AMD Ryzen 3900X CPU. Images were all rendered at a resolution of $512 \times 512$ during training.

In our tests, we use two different types of data sets.
First, it is evaluated how DiffTetVR fares at reconstructing preshaded volumes from a set of rendered scientific volume data set images and associated camera poses. To mirror the experiments by Weiss and Westermann~\cite{DiffDVR}, we use the Tooth data set from the \href{http://klacansky.com/open-scivis-datasets/}{Open SciVis Datasets} and apply the same "tooth3gauss" transfer function, which was obtained from \url{https://github.com/shamanDevel/DiffDVR}. Secondly, we show the general correctness of the technique by applying it to some of the synthetic surface mesh test data sets provided by Mildenhall~\etal~\cite{NeRF}. We use the data set \href{https://www.blendswap.com/blend/23962}{Hotdog by erickfree} (CC-0) as an example for a scene with hard surfaces.

\subsection{Color-only optimization}\label{sec:results-color}

In \cref{fig:tooth-img}, it can be seen that DiffTetVR can successfully reconstruct vertex colors through inverse rendering for the volumetric \textit{Tooth} data set using the same transfer function as employed by Weiss and Westermann~\cite{DiffDVR}. Some differences between the ground truth and reconstruction are visible and can be attributed to interpolation errors. While for the ground truth image, scalar densities are interpolated before applying the color and opacity transfer function, reconstructing a pre-shaded volume means that DiffTetVR directly interpolates the vertex colors instead of the scalar quantity. To avoid these issues, the code of DiffTetVR would simply need to be adapted to not only support reconstruction of pre-shaded volumes, but also scalar density volumes and transfer functions. On the test system, convergence of the optimization process was achieved in about 3 seconds even when using a constant vertex color learning rate of $0.08$. Further speedup can be achieved by starting with a higher learning rate and using consecutive learning rate decay. This shows the great speed of the Vulkan renderer detailed in \cref{sec:impl}.

In \cref{fig:hotdog-img}, we see that DiffTetVR also reconstructs surface renderings at an acceptable quality level. Here, two weaknesses of the technique can be seen. Firstly, DiffTetVR employs a volumetric emission-absorption model, which is unnecessary for reconstructing surface meshes without any volumetric component. In such a case, one would usually prefer to reconstruct a triangle mesh, and it may be more meaningful to use marching tetrahedra-based techniques like DMTet~\cite{DMTet} when no volumetric component needs to be reconstructed. Secondly, while a triangle mesh can be extracted from the volumetric tetrahedra, a lot of resolution is wasted in empty regions. A solution for this that increases performance and reduces memory consumption could be to remove tetrahedra that are still empty after a few training epochs.

\begin{figure*}[t]
\centering
\begin{minipage}{6in}
  \centering
\raisebox{-0.5\height}{\includegraphics[width=0.324\linewidth]{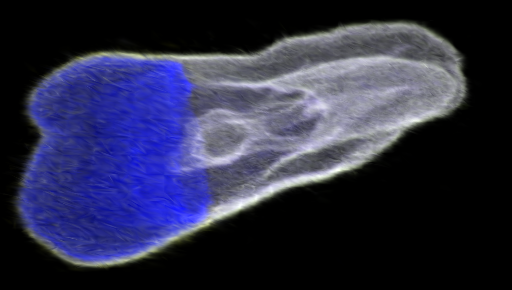}\hspace{0.5cm}}\hspace{1cm}
  \raisebox{-0.5\height}{\includegraphics[width=0.48\linewidth]{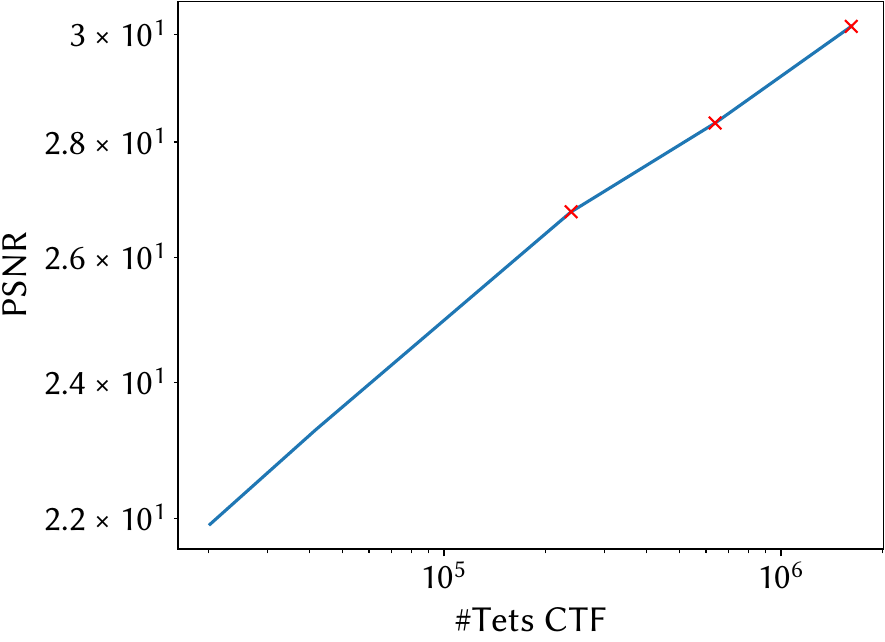}}
\end{minipage}
\caption{Reconstruction of \textit{Tooth} data set using adaptive subdivision as described in \cref{sec:results-subdiv}. While optimization results are better than using vertex position optimization without subdivision (cf.~\cref{fig:tooth-dia}
top right), final PSNR results cannot reach color-only optimization quality as shown in \cref{fig:tooth-img}~middle and \cref{fig:tooth-dia}~top left. Hyperparameter choices can be found \href{https://github.com/chrismile/DiffTetVR/blob/0643199f745c98895ae973b40a99ee7462d945b3/pytests/run.py\#L175}{online}.}
\label{fig:tooth-adaptive}
\end{figure*}

\begin{figure*}[t]
\centering
\includegraphics[width=0.45\linewidth]{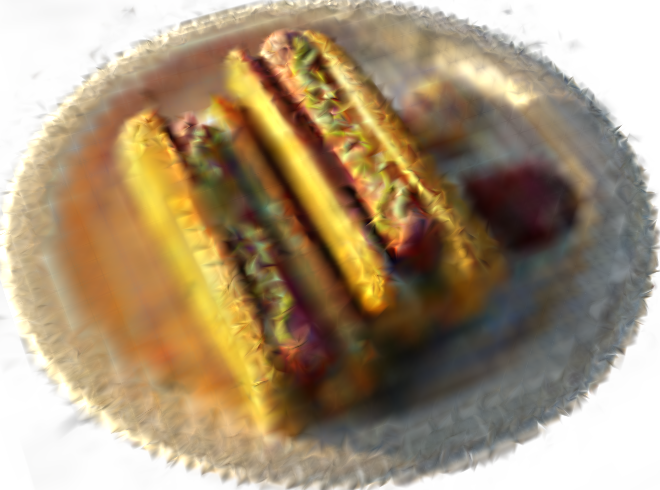}\hspace{0.5cm}
\caption{Reconstruction of \textit{Hotdog} data set using adaptive subdivision. Vertex position optimization has been disabled. Test set PSNR of 24.4 is lower than using color-only optimization in \cref{fig:hotdog-img} with PSNR of 25.2. Hyperparameter choices can be found \href{https://github.com/chrismile/DiffTetVR/blob/0643199f745c98895ae973b40a99ee7462d945b3/pytests/run.py\#L250}{online}.}
\label{fig:hotdog-adaptive}
\end{figure*}

\subsection{Adaptive subdivision}\label{sec:results-subdiv}

The conclusion of the last section about unneeded empty tetrahedra directly raises the question whether adaptive subdivision paired with vertex position optimization and a tetrahedral shape regularizer can solve this concern. Unfortunately, our experiments have shown that this does not work well as it is currently implemented in DiffTetVR.

In a first experiment, we have tested what happens when enabling the optimization of vertex positions in addition to the optimization of vertex colors for the \textit{Tooth} data set. In different experiments, we have checked how concergence changes with different settings of the vertex position learning rate $\text{lr}_p$, the regularization strength $\lambda$ and the softplus parameter $\beta$. When fixing $\beta=100$ and $\lambda=10$, it can be seen in \cref{fig:tooth-dia} that the optimization process is only stable (i.e., does not generate degenerate tetrahedra) for the lowest learning rates of 1e-6 and 5e-6. But even then, the PSNR of approx.\ 26.9 is significantly lower than the peak of 33.5 that can be seen for color-only optimization. This raises the concern whether the local optimization of vertex positions by backpropagating color image differences through the rendering process and barycentric interpolation can somehow lead to a better global optimum. Our results do not indicate that this is the case, at least not how it is currently implemented.

In a second experiment, we enabled adaptive refinement for the tetrahedral mesh. The training process is conducted as follows. One epoch is defined as training with 800 \textit{Tooth} data set renderings. For one epoch, only vertex colors are optimized. Then, for one epoch, the vertex colors and positions are jointly optimized. For the next epoch, absolute color gradients are accumulated at every vertex without doing optimization steps. Finally, the $5\%$ vertices with the highest absolute color gradients are taken and the incident tetrahedra are subdivided as demonstrated in \cref{sec:tet-subdiv}. This process is repeated until a target amount of tetrahedra is reached. In the last iteration, only vertex colors are optimized for the final result.

However, results with adaptive subdivision were also not producing satisfactory results (cf.~\cref{fig:tooth-adaptive}). Even when setting the vertex position learning rate to 0, the local error did not decrease. This is potentially due to the interpolation error of the prolongated tetrahedra generated during subdivision (compare \cref{fig:hotdog-adaptive}). And as mentioned in the last section, we have not been able to stabilize the training process when using vertex position learning rates not equal to zero.

\section{Discussion and Conclusion}\label{sec:discussion}

The greatest strength of DiffTetVR is the generalization of DiffDVR by Weiss and Westermann~\cite{DiffDVR} from regular grids to tetrahedral meshes. This allows mesh refinement only close to the surface to assign high mesh resolution to where it is truly necessary, thus potentially avoiding high tessellation in empty regions.

Unfortunately, the optimization of vertex positions has proven to be quite fickle. Below, a summary of the largest weaknesses of the presented approach can be found.

\begin{itemize}
    \item The optimization of vertex positions using the presented technique has proven to be quite unstable. This gets worse the more subdivided the mesh becomes. As tetrahedra get smaller, the chance of inversion increases. Re-tesselation after each step as done by other works like Govindarajan~\etal~\cite{RadiantFoam} has the advantage of avoiding these issues and may turn out to be the more stable approach.
    \item Subdivision using the technique described in \cref{sec:tet-subdiv} does not seem to sufficiently lower the local error, so the same region of the mesh gets recursively subdivided more and more. After too many optimization epochs, solving the CSPs and updating the OpenVolumeMesh data structure dominates computation time and the optimization progress slowly comes to a halt.
    \item Unlike classical NeRF approaches, DiffTetVR, just like DiffDVR~\cite{DiffDVR}, does not model any directional lighting effects, nor does it model volume scattering like the work by Leonard~\etal~\cite{leonard2025lighttransportawarediffusionposterior}.
    \item Real world applications often prefer triangle meshes, and thus it can be better to directly use marching tetrahedra-based techniques like DMTet~\cite{DMTet} instead of employing volume rendering approaches.
\end{itemize}

We encourage readers to take inspiration from our ideas and to reuse code that was written as part of this work to build future works that are able to achieve the goal DiffTetVR set out to solve and enable adaptive reconstruction of volumetric data from images.

\newpage
\section{Appendix: Painter's algorithm}\label{sec:interlock}

As discussed in the main manuscript, a topological sorting pass is necessary for implementing the Painter's algorithm~\cite{deBerg1993,deBerg2008}. While the directed acyclic graph (DAG) described by Williams~\cite{WilliamsMeshSortDAG} could be constructed efficiently on a GPU, to our knowledge no efficient parallel DAG sorter implementation currently exists for GPUs.

The VTK toolkit~\cite{vtkBook} provides an implementation of projected tetrahedra (PT)~\cite{ShirleyTuchmanDecomp} based on sorting by the depth of the centroid of the tetrahedral elements. While this can lead to incorrect depth order and rendering artifacts, in many cases this approximation provides sufficient quality. While VTK implements the generation of the projected tetrahedra triangles on the CPU, we have ported their code to a compute shader in Vulkan. Generated triangles are appended linearly to a large buffer in GPU memory by using a global atomic triangle counter variable. We make use of the Vulkan radix sort implementation provided by the Fuchsia project \cite{FuchsiaRadixSort} for fast sorting of the triangles on the GPU.

One more challenge for a Painter's algorithm-based differentiable renderer besides topological sorting is the backward pass.
While it is easy to just rasterize the tetrahedra in inverse depth order in the backward pass, one needs access to intermediate values and gradients during the backpropagation process (cf.~\cref{sec:bckwd}).

To solve this issue, we propose the use of a GPU feature called fragment shader interlock (Vulkan and OpenGL) \cite{FragmentShaderInterlock} and raster order views (Direct3D). Using this feature, we can load and store the intermediate values and gradients from and to a per-pixel data structure in global GPU memory, and provide critical sections in the shader that ensure that memory accesses are linearized for the same pixel. This way, race conditions can be avoided.

\begin{figure*}[t]
\centering
\includegraphics[width=0.95\linewidth]{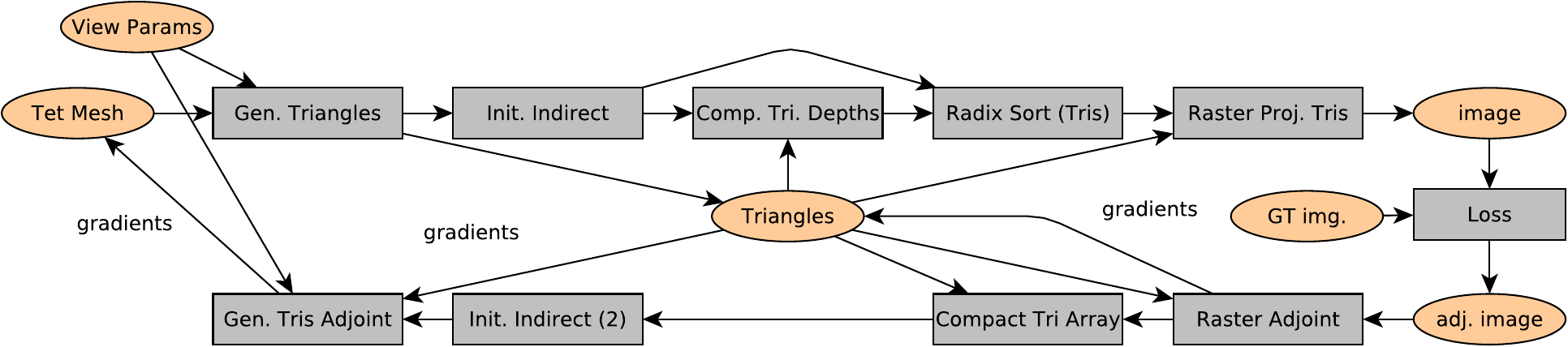}
\caption{Execution pipeline for the differentiable projected triangles implementation. After generating the projected triangles using the technique by Shirley and Tuchman~\cite{ShirleyTuchmanDecomp}, indirect command buffers are generated for subsequent passes, as the number of execution threads depends on the number of generated triangles. Next, depths are generated for all triangles and they are sorted using a GPU-based radix sort algorithm. Finally, the projected triangles are rasterized in sorted order and an image is generated. The loss between the generated image and a ground truth image is generated. In the backward pass, by differentiating the loss, an adjoint image is generated. An adjoint raster pass uses the inversion trick by Weiß and Westermann~\cite{DiffDVR} and a backward pass for the barycentric interpolation to compute gradients for the properties of the projected tetrahedra triangles. A compaction pass computes which tetrahedra have resulted in visible triangles and computes a map from the tetrahedron index to the index of the first triangle. Again, an indirect command buffer is generated dependent on the number of visible tetrahedra. Finally, an adjoint triangle projection pass uses the chain rule to transfer projected tetrahedra triangle gradients to world space tetrahedra gradients.}
\label{fig:pt-pipeline}
\end{figure*}

In our source code, we provide an alternative implementation using fragment shader interlock that currently uses a naive sort by the center point depth of the tets. The execution pipeline is shown in \cref{fig:pt-pipeline}. If an efficient parallel DAG sorter should ever become possible to implement on a GPU, the sorter can be replaced and this renderer can be used as a potentially even more performant alternative to our proposed method using per-pixel linked lists~\cite{Yang2010}. Per-pixel linked list also suffer from unbounded memory requirements, while PT can work with a fixed amount of memory.

\newpage
\section{Appendix: Forward rendering}\label{sec:fwd-exact}

In this section, the forward rendering equations are given for the case $\alpha_0 \ne \alpha_1$ for \cref{eq:accum}.

\begin{equation}
a(\alpha_0, \alpha_1) = \frac{\alpha_0 - \alpha_1}{2}
\end{equation}

We will assume that $a(\alpha_0, \alpha_1) \ne 0$, as otherwise $\alpha_0 = \alpha_1$, and the simplified formulas derived in the last sections could be used. We use the helper function below to simplify the formulas.

\begin{equation}
b(\alpha_0, \alpha_1) = \frac{\alpha_0}{\alpha_1 - \alpha_0}
\end{equation}

In the following equations, the arguments of the used auxiliary functions will be omitted for the sake of readability.

\begin{equation}
G(t, c_0, c_1, \alpha_0, \alpha_1) = \begin{cases}
    \erfi(\sqrt{a}(t + b)),& \text{if } a\ge 0\\
    \erf(\sqrt{-a}(t + b)),& \text{else}
\end{cases}
\end{equation}

\begin{equation}
H(t, c_0, c_1, \alpha_0, \alpha_1) = \begin{cases}
    \erfi(\sqrt{a}b),& \text{if } a\ge 0\\
    \erf(\sqrt{-a}b),& \text{else}
\end{cases}
\end{equation}

\begin{equation}
p_0(c_0, \alpha_0) = \alpha_0 c_0
\end{equation}

\begin{equation}
p_1(c_0, c_1, \alpha_0, \alpha_1) = -2 \alpha_0 c_0 + \alpha_1 c_0 + \alpha_0 c_1
\end{equation}

\begin{equation}
p_2(c_0, c_1, \alpha_0, \alpha_1) = \alpha_0 c_0 - \alpha_1 c_0 - \alpha_0 c_1 + \alpha_1 c_1
\end{equation}

\begin{equation}
A(t, c_0, c_1, \alpha_0, \alpha_1) = \frac{\sqrt{\pi}(G - H)(p_0 - b p_1 + b^2 p_2)}{2\sqrt{a}}
\end{equation}

\begin{equation}
B(t, c_0, c_1, \alpha_0, \alpha_1) = \frac{(p_1 - 2 b p_2)(e^{a(t + b)^2} - e^{ab^2})}{2 a}
\end{equation}

\begin{equation}
C(t, c_0, c_1, \alpha_0, \alpha_1) = \frac{p_2\left((t + b)e^{a(t + b)^2} - b e^{ab^2} + \frac{\sqrt{\pi} (H - G)}{2 \sqrt{|a|}}\right)}{2 a}
\end{equation}

\begin{equation}
c_{acc}(t, c_0, c_1, \alpha_0, \alpha_1) = e^{a} * (A + B + C)
\end{equation}

\begin{equation}
\alpha_{acc}(t, c_0, c_1, \alpha_0, \alpha_1) = 1 - e^{-\alpha_0t + at^2}
\end{equation}

Other formulas:

\begin{equation}
\erfi(z) = -i \erf(iz)
\end{equation}

These formulas are comparable to the ones derived by Williams and Max~\cite{TetIntegration} for the exact accumulation of color along a ray in a tetrahedral element. In their work they state that ``it remains to be seen if these expressions can be evaluated so as to permit interactive rendering''~\cite{TetIntegration}. As mentioned in \cref{sec:fwd}, using this formula has proven to be infeasible due to the numeric instability of computing the difference of the $\erfi$ functions used in the auxiliary terms $G$ and $H$.

\newpage
\section{Appendix: Backward rendering constant per-cell color}\label{sec:bckwd-const}

\begin{equation}
c_{acc}(t, c, \alpha) = \left( 1 - e^{-\alpha t} \right) c
\end{equation}
\begin{equation}
a_{acc}(t, \alpha) = 1 - e^{-\alpha t}
\end{equation}
\begin{equation}
\frac{\partial c_{acc}(t, c, \alpha)}{\partial c} = 1 - e^{-\alpha t}
\end{equation}
\begin{equation}
\frac{\partial c_{acc}(t, c, \alpha)}{\partial \alpha} = t e^{-\alpha t} c
\end{equation}
\begin{equation}
\frac{\partial c_{acc}(t, c, \alpha)}{\partial t} = \alpha e^{-\alpha t} c
\end{equation}
\begin{equation}
\frac{\partial a_{acc}(t, \alpha)}{\partial \alpha} = t e^{-\alpha t}
\end{equation}
\begin{equation}
\frac{\partial a_{acc}(t, \alpha)}{\partial t} = \alpha e^{-\alpha t}
\end{equation}

\newpage
\bibliographystyle{IEEEtran}
\bibliography{IEEEabrv,main}

\begin{thebibliography}{10}
\providecommand{\url}[1]{#1}
\csname url@samestyle\endcsname
\providecommand{\newblock}{\relax}
\providecommand{\bibinfo}[2]{#2}
\providecommand{\BIBentrySTDinterwordspacing}{\spaceskip=0pt\relax}
\providecommand{\BIBentryALTinterwordstretchfactor}{4}
\providecommand{\BIBentryALTinterwordspacing}{\spaceskip=\fontdimen2\font plus
\BIBentryALTinterwordstretchfactor\fontdimen3\font minus
  \fontdimen4\font\relax}
\providecommand{\BIBforeignlanguage}[2]{{%
\expandafter\ifx\csname l@#1\endcsname\relax
\typeout{** WARNING: IEEEtran.bst: No hyphenation pattern has been}%
\typeout{** loaded for the language `#1'. Using the pattern for}%
\typeout{** the default language instead.}%
\else
\language=\csname l@#1\endcsname
\fi
#2}}
\providecommand{\BIBdecl}{\relax}
\BIBdecl

\bibitem{DiffRenderSurvey}
\BIBentryALTinterwordspacing
H.~Kato, D.~Beker, M.~Morariu, T.~Ando, T.~Matsuoka, W.~Kehl, and A.~Gaidon,
  ``Differentiable rendering: A survey,'' 2020. [Online]. Available:
  \url{https://arxiv.org/abs/2006.12057}
\BIBentrySTDinterwordspacing

\bibitem{SoftRas}
\BIBentryALTinterwordspacing
S.~Liu, T.~Li, W.~Chen, and H.~Li, ``Soft rasterizer: A differentiable renderer
  for image-based 3d reasoning,'' 2019. [Online]. Available:
  \url{https://arxiv.org/abs/1904.01786}
\BIBentrySTDinterwordspacing

\bibitem{NvDiffRast}
\BIBentryALTinterwordspacing
S.~Laine, J.~Hellsten, T.~Karras, Y.~Seol, J.~Lehtinen, and T.~Aila, ``Modular
  primitives for high-performance differentiable rendering,'' \emph{ACM Trans.
  Graph.}, vol.~39, no.~6, Nov. 2020. [Online]. Available:
  \url{https://doi.org/10.1145/3414685.3417861}
\BIBentrySTDinterwordspacing

\bibitem{LinPointCloud}
\BIBentryALTinterwordspacing
C.-H. Lin, C.~Kong, and S.~Lucey, ``Learning efficient point cloud generation
  for dense 3d object reconstruction,'' 2017. [Online]. Available:
  \url{https://arxiv.org/abs/1706.07036}
\BIBentrySTDinterwordspacing

\bibitem{Kerbl3DGS}
\BIBentryALTinterwordspacing
B.~Kerbl, G.~Kopanas, T.~Leimkühler, and G.~Drettakis, ``{3D} gaussian
  splatting for real-time radiance field rendering,'' 2023. [Online].
  Available: \url{https://arxiv.org/abs/2308.04079}
\BIBentrySTDinterwordspacing

\bibitem{DiffDVR}
\BIBentryALTinterwordspacing
S.~Weiss and R.~Westermann, ``Differentiable direct volume rendering,''
  \emph{IEEE Transactions on Visualization and Computer Graphics}, vol.~28,
  no.~1, pp. 562--572, 2022. [Online]. Available:
  \url{https://doi.org/10.1109/TVCG.2021.3114769}
\BIBentrySTDinterwordspacing

\bibitem{PorterDuff}
\BIBentryALTinterwordspacing
T.~Porter and T.~Duff, ``Compositing digital images,'' \emph{SIGGRAPH Comput.
  Graph.}, vol.~18, no.~3, pp. 253--259, Jan. 1984. [Online]. Available:
  \url{https://doi.org/10.1145/964965.808606}
\BIBentrySTDinterwordspacing

\bibitem{Mitsuba2}
\BIBentryALTinterwordspacing
M.~Nimier-David, D.~Vicini, T.~Zeltner, and W.~Jakob, ``Mitsuba 2: a
  retargetable forward and inverse renderer,'' \emph{ACM Trans. Graph.},
  vol.~38, no.~6, Nov. 2019. [Online]. Available:
  \url{https://doi.org/10.1145/3355089.3356498}
\BIBentrySTDinterwordspacing

\bibitem{Yang2010}
\BIBentryALTinterwordspacing
J.~C. Yang, J.~Hensley, H.~Gr\"{u}n, and N.~Thibieroz, ``Real-time concurrent
  linked list construction on the gpu,'' in \emph{Proceedings of the 21st
  Eurographics Conference on Rendering}, ser. EGSR'10.\hskip 1em plus 0.5em
  minus 0.4em\relax Goslar, DEU: Eurographics Association, 2010, pp.
  1297--1304. [Online]. Available:
  \url{https://doi.org/10.1111/j.1467-8659.2010.01725.x}
\BIBentrySTDinterwordspacing

\bibitem{nimierdavid2022unbiased}
\BIBentryALTinterwordspacing
M.~Nimier-David, T.~M\"uller, A.~Keller, and W.~Jakob, ``Unbiased inverse
  volume rendering with differential trackers,'' \emph{ACM Trans. Graph.},
  vol.~41, no.~4, pp. 44:1--44:20, Jul. 2022. [Online]. Available:
  \url{https://doi.org/10.1145/3528223.3530073}
\BIBentrySTDinterwordspacing

\bibitem{novak14residual}
\BIBentryALTinterwordspacing
J.~Nov\'ak, A.~Selle, and W.~Jarosz, ``Residual ratio tracking for estimating
  attenuation in participating media,'' \emph{ACM Transactions on Graphics
  (Proceedings of SIGGRAPH Asia)}, vol.~33, no.~6, Nov. 2014. [Online].
  Available: \url{https://dl.acm.org/doi/10.1145/2661229.2661292}
\BIBentrySTDinterwordspacing

\bibitem{leonard2025lighttransportawarediffusionposterior}
\BIBentryALTinterwordspacing
L.~Leonard, N.~Thuerey, and R.~Westermann, ``Light transport-aware diffusion
  posterior sampling for single-view reconstruction of 3d volumes,'' 2025.
  [Online]. Available: \url{https://arxiv.org/abs/2501.05226}
\BIBentrySTDinterwordspacing

\bibitem{DMTet}
\BIBentryALTinterwordspacing
T.~Shen, J.~Gao, K.~Yin, M.-Y. Liu, and S.~Fidler, ``Deep marching tetrahedra:
  a hybrid representation for high-resolution 3d shape synthesis,'' in
  \emph{Advances in Neural Information Processing Systems (NeurIPS)}, 2021.
  [Online]. Available: \url{https://research.nvidia.com/labs/toronto-ai/DMTet/}
\BIBentrySTDinterwordspacing

\bibitem{MarchingTet}
\BIBentryALTinterwordspacing
A.~Doi and A.~Koide, ``An efficient method of triangulating equi-valued
  surfaces by using tetrahedral cells,'' \emph{IEICE Transactions on
  Information and Systems}, vol.~74, pp. 214--224, 1991. [Online]. Available:
  \url{https://search.ieice.org/bin/summary.php?id=e74-d_1_214}
\BIBentrySTDinterwordspacing

\bibitem{yu2024gaussianopacityfieldsefficient}
\BIBentryALTinterwordspacing
Z.~Yu, T.~Sattler, and A.~Geiger, ``Gaussian opacity fields: Efficient adaptive
  surface reconstruction in unbounded scenes,'' 2024. [Online]. Available:
  \url{https://arxiv.org/abs/2404.10772}
\BIBentrySTDinterwordspacing

\bibitem{TetraNeRF}
\BIBentryALTinterwordspacing
J.~Kulhanek and T.~Sattler, ``Tetra-nerf: Representing neural radiance fields
  using tetrahedra,'' 2023. [Online]. Available:
  \url{https://arxiv.org/abs/2304.09987}
\BIBentrySTDinterwordspacing

\bibitem{NeRF}
\BIBentryALTinterwordspacing
B.~Mildenhall, P.~P. Srinivasan, M.~Tancik, J.~T. Barron, R.~Ramamoorthi, and
  R.~Ng, ``Nerf: representing scenes as neural radiance fields for view
  synthesis,'' \emph{Commun. ACM}, vol.~65, no.~1, p. 99–106, Dec. 2021.
  [Online]. Available: \url{https://doi.org/10.1145/3503250}
\BIBentrySTDinterwordspacing

\bibitem{TetSplatting}
\BIBentryALTinterwordspacing
C.~Gu, Z.~Yang, Z.~Pan, X.~Zhu, and L.~Zhang, ``Tetrahedron splatting for 3d
  generation,'' 2024. [Online]. Available:
  \url{https://arxiv.org/abs/2406.01579}
\BIBentrySTDinterwordspacing

\bibitem{RadiantFoam}
\BIBentryALTinterwordspacing
S.~Govindarajan, D.~Rebain, K.~M. Yi, and A.~Tagliasacchi, ``Radiant foam:
  Real-time differentiable ray tracing,'' 2025. [Online]. Available:
  \url{https://arxiv.org/abs/2502.01157}
\BIBentrySTDinterwordspacing

\bibitem{TetIntegration}
\BIBentryALTinterwordspacing
P.~L. Williams and N.~Max, ``A volume density optical model,'' in
  \emph{Proceedings of the 1992 Workshop on Volume Visualization}, ser. VVS
  '92.\hskip 1em plus 0.5em minus 0.4em\relax New York, NY, USA: Association
  for Computing Machinery, 1992, p. 61–68. [Online]. Available:
  \url{https://doi.org/10.1145/147130.147151}
\BIBentrySTDinterwordspacing

\bibitem{TetMeshRTShellTraversal}
\BIBentryALTinterwordspacing
A.~Şahıstan, S.~Demirci, N.~Morrical, S.~Zellmann, A.~Aman, I.~Wald, and
  U.~Güdükbay, ``Ray-traced shell traversal of tetrahedral meshes for direct
  volume visualization,'' in \emph{2021 IEEE Visualization Conference (VIS)},
  2021, pp. 91--95. [Online]. Available:
  \url{https://doi.org/10.1109/VIS49827.2021.9623298}
\BIBentrySTDinterwordspacing

\bibitem{UnstructuredMeshSpaceSkip}
\BIBentryALTinterwordspacing
N.~Morrical, W.~Usher, I.~Wald, and V.~Pascucci, ``Efficient space skipping and
  adaptive sampling of unstructured volumes using hardware accelerated ray
  tracing,'' in \emph{2019 IEEE Visualization Conference (VIS)}, 2019, pp.
  256--260. [Online]. Available:
  \url{https://doi.org/10.1109/VISUAL.2019.8933539}
\BIBentrySTDinterwordspacing

\bibitem{TetMeshPointLocationRTX}
\BIBentryALTinterwordspacing
I.~Wald, W.~Usher, N.~Morrical, L.~Lediaev, and V.~Pascucci, ``Rtx beyond ray
  tracing: exploring the use of hardware ray tracing cores for tet-mesh point
  location,'' in \emph{Proceedings of the Conference on High-Performance
  Graphics}, ser. HPG '19.\hskip 1em plus 0.5em minus 0.4em\relax Goslar, DEU:
  Eurographics Association, 2022, p. 7–13. [Online]. Available:
  \url{https://doi.org/10.2312/hpg.20191189}
\BIBentrySTDinterwordspacing

\bibitem{deBerg1993}
\BIBentryALTinterwordspacing
M.~de~Berg, \emph{Depth orders in three dimensions}.\hskip 1em plus 0.5em minus
  0.4em\relax Berlin, Heidelberg: Springer Berlin Heidelberg, 1993, pp.
  135--144. [Online]. Available: \url{https://doi.org/10.1007/BFb0029824}
\BIBentrySTDinterwordspacing

\bibitem{deBerg2008}
\BIBentryALTinterwordspacing
M.~de~Berg, O.~Cheong, M.~van Krefeld, and M.~Overmars, \emph{Binary Space
  Partitions}.\hskip 1em plus 0.5em minus 0.4em\relax Berlin, Heidelberg:
  Springer Berlin Heidelberg, 2008, pp. 259--281. [Online]. Available:
  \url{https://doi.org/10.1007/978-3-540-77974-2_12}
\BIBentrySTDinterwordspacing

\bibitem{DepthPresortedTris}
\BIBentryALTinterwordspacing
G.~Chen, P.~V. Sander, D.~Nehab, L.~Yang, and L.~Hu, ``Depth-presorted triangle
  lists,'' \emph{ACM Trans. Graph.}, vol.~31, no.~6, Nov. 2012. [Online].
  Available: \url{https://doi.org/10.1145/2366145.2366179}
\BIBentrySTDinterwordspacing

\bibitem{WilliamsMeshSortDAG}
\BIBentryALTinterwordspacing
P.~L. Williams, ``Visibility-ordering meshed polyhedra,'' \emph{ACM Trans.
  Graph.}, vol.~11, no.~2, p. 103–126, Apr. 1992. [Online]. Available:
  \url{https://doi.org/10.1145/130826.130899}
\BIBentrySTDinterwordspacing

\bibitem{NelsonHanrahanCrawfisSort}
\BIBentryALTinterwordspacing
N.~Max, P.~Hanrahan, and R.~Crawfis, ``Area and volume coherence for efficient
  visualization of 3d scalar functions,'' \emph{SIGGRAPH Comput. Graph.},
  vol.~24, no.~5, p. 27–33, Nov. 1990. [Online]. Available:
  \url{https://doi.org/10.1145/99308.99315}
\BIBentrySTDinterwordspacing

\bibitem{GPUParTopSort}
R.~Saxena, M.~Jain, and D.~P. Sharma, ``Gpu-based parallelization of
  topological sorting,'' in \emph{Proceedings of First International Conference
  on Smart System, Innovations and Computing}, A.~K. Somani, S.~Srivastava,
  A.~Mundra, and S.~Rawat, Eds.\hskip 1em plus 0.5em minus 0.4em\relax
  Singapore: Springer Singapore, 2018, pp. 411--421.

\bibitem{ShirleyTuchmanDecomp}
\BIBentryALTinterwordspacing
P.~Shirley and A.~Tuchman, ``A polygonal approximation to direct scalar volume
  rendering,'' \emph{SIGGRAPH Comput. Graph.}, vol.~24, no.~5, p. 63–70, nov
  1990. [Online]. Available: \url{https://doi.org/10.1145/99308.99322}
\BIBentrySTDinterwordspacing

\bibitem{ProjectingTetrahedra}
M.~Kraus, W.~Qiao, and D.~Ebert, ``Projecting tetrahedra without rendering
  artifacts,'' in \emph{IEEE Visualization 2004}, 2004, pp. 27--34.

\bibitem{SteinExp}
\BIBentryALTinterwordspacing
C.~M. Stein, B.~G. Becker, and N.~L. Max, ``Sorting and hardware assisted
  rendering for volume visualization,'' in \emph{Proceedings of the 1994
  Symposium on Volume Visualization}, ser. VVS '94.\hskip 1em plus 0.5em minus
  0.4em\relax New York, NY, USA: Association for Computing Machinery, 1994, p.
  83–89. [Online]. Available: \url{https://doi.org/10.1145/197938.197971}
\BIBentrySTDinterwordspacing

\bibitem{vtkBook}
W.~Schroeder, K.~Martin, and B.~Lorensen, \emph{The Visualization Toolkit (4th
  ed.)}.\hskip 1em plus 0.5em minus 0.4em\relax Kitware, 2006.

\bibitem{ABuffer}
\BIBentryALTinterwordspacing
L.~Carpenter, ``The a -buffer, an antialiased hidden surface method,'' in
  \emph{Proceedings of the 11th Annual Conference on Computer Graphics and
  Interactive Techniques}, ser. SIGGRAPH '84.\hskip 1em plus 0.5em minus
  0.4em\relax New York, NY, USA: Association for Computing Machinery, 1984, p.
  103–108. [Online]. Available: \url{https://doi.org/10.1145/800031.808585}
\BIBentrySTDinterwordspacing

\bibitem{Everitt2001}
\BIBentryALTinterwordspacing
C.~Everitt, ``Interactive order-independent transparency,'' \emph{NVIDIA
  Corporation}, Oct. 2001. [Online]. Available:
  \url{https://my.eng.utah.edu/~cs5610/handouts/order_independent_transparency.pdf}
\BIBentrySTDinterwordspacing

\bibitem{Bavoil2008}
L.~Bavoil and K.~Myers, ``Order independent transparency with dual depth
  peeling,'' Jan. 2008.

\bibitem{Salvi2014}
M.~Salvi and K.~Vaidyanathan, ``Multi-layer alpha blending,'' in
  \emph{Proceedings of the 18th Meeting of the ACM SIGGRAPH Symposium on
  Interactive 3D Graphics and Games}, ser. I3D '14.\hskip 1em plus 0.5em minus
  0.4em\relax New York, NY, USA: ACM, 2014, pp. 151--158.

\bibitem{Muenstermann2018}
\BIBentryALTinterwordspacing
C.~M\"{u}nstermann, S.~Krumpen, R.~Klein, and C.~Peters, ``Moment-based
  order-independent transparency,'' \emph{Proc. ACM Comput. Graph. Interact.
  Tech.}, vol.~1, no.~1, Jul. 2018. [Online]. Available:
  \url{https://doi.org/10.1145/3203206}
\BIBentrySTDinterwordspacing

\bibitem{StochasticTransparency}
E.~Enderton, E.~Sintorn, P.~Shirley, and D.~Luebke, ``Stochastic
  transparency,'' in \emph{I3D '10: Proceedings of the 2010 symposium on
  Interactive 3D graphics and games}, New York, NY, USA, 2010, pp. 157--164.

\bibitem{schopf2014burmann}
\BIBentryALTinterwordspacing
H.~M. Sch{\"o}pf and P.~H. Supancic, ``On {B{\"u}rmann’s} theorem and its
  application to problems of linear and nonlinear heat transfer and
  diffusion,'' \emph{The Mathematica Journal}, vol.~16, no.~11, 2014. [Online].
  Available: \url{https://doi.org/10.3888%2Ftmj.16-11}
\BIBentrySTDinterwordspacing

\bibitem{TransparentLines}
M.~Kern, C.~Neuhauser, T.~Maack, M.~Han, W.~Usher, and R.~Westermann, ``A
  comparison of rendering techniques for 3d line sets with transparency,''
  \emph{IEEE Transactions on Visualization and Computer Graphics}, vol.~27,
  no.~8, pp. 3361--3376, 2021.

\bibitem{VulkanSpec}
\BIBentryALTinterwordspacing
{The Khronos Vulkan Working Group}. (2024) {Vulkan 1.3.291 - A Specification}.
  [Online]. Available:
  \url{https://registry.khronos.org/vulkan/specs/1.3-extensions/html/vkspec.html}
\BIBentrySTDinterwordspacing

\bibitem{OpenVolumeMeshPaper}
\BIBentryALTinterwordspacing
M.~Kremer, D.~Bommes, and L.~Kobbelt, ``Openvolumemesh -- a versatile
  index-based data structure for 3d polytopal complexes,'' in \emph{Proceedings
  of the 21st International Meshing Roundtable}, X.~Jiao and J.-C. Weill,
  Eds.\hskip 1em plus 0.5em minus 0.4em\relax Berlin, Heidelberg: Springer
  Berlin Heidelberg, 2013, pp. 531--548. [Online]. Available:
  \url{https://www.graphics.rwth-aachen.de/media/papers/MKremer_OVM.pdf}
\BIBentrySTDinterwordspacing

\bibitem{AdaptiveThinShellTetMesh}
K.~Erleben, H.~Dohlmann, and J.~Sporring, ``The adaptive thin shell tetrahedral
  mesh,'' in \emph{Journal of WSCG}, 01 2005, pp. 17--24.

\bibitem{SympyPaper}
\BIBentryALTinterwordspacing
A.~Meurer, C.~P. Smith, M.~Paprocki, O.~\v{C}ert\'{i}k, S.~B. Kirpichev,
  M.~Rocklin, A.~Kumar, S.~Ivanov, J.~K. Moore, S.~Singh, T.~Rathnayake,
  S.~Vig, B.~E. Granger, R.~P. Muller, F.~Bonazzi, H.~Gupta, S.~Vats,
  F.~Johansson, F.~Pedregosa, M.~J. Curry, A.~R. Terrel, v.~Rou\v{c}ka,
  A.~Saboo, I.~Fernando, S.~Kulal, R.~Cimrman, and A.~Scopatz, ``{SymPy}:
  Symbolic computing in {Python},'' \emph{PeerJ Computer Science}, vol.~3, p.
  e103, Jan. 2017. [Online]. Available:
  \url{https://doi.org/10.7717/peerj-cs.103}
\BIBentrySTDinterwordspacing

\bibitem{TetQualityShort}
\BIBentryALTinterwordspacing
J.~R. Shewchuk, ``What is a good linear element? {Interpolation}, conditioning,
  and quality measures,'' in \emph{Proceedings of the 11th International
  Meshing Roundtable, {IMR} 2002, Ithaca, New York, USA, September 15-18,
  2002}, 2002, pp. 115--126. [Online]. Available:
  \url{http://imr.sandia.gov/papers/abstracts/Sh247.html}
\BIBentrySTDinterwordspacing

\bibitem{TetQualityFull}
\BIBentryALTinterwordspacing
------, ``What is a good linear finite element? {Interpolation}, conditioning,
  anisotropy, and quality measures,'' 2002, unpublished preprint. [Online].
  Available: \url{http://www.cs.berkeley.edu/~jrs/papers/elemj.pdf}
\BIBentrySTDinterwordspacing

\bibitem{Parthasarathy}
V.~N. Parthasarathy, C.~M. Graichen, and A.~Hathaway, ``Fast evaluation \&
  improvement of tetrahedral 3-d grid quality,'' 1991.

\bibitem{AbsGS}
\BIBentryALTinterwordspacing
Z.~Ye, W.~Li, S.~Liu, P.~Qiao, and Y.~Dou, ``Absgs: Recovering fine details for
  3d gaussian splatting,'' 2024. [Online]. Available:
  \url{https://arxiv.org/abs/2404.10484}
\BIBentrySTDinterwordspacing

\bibitem{schoenberger2016sfm}
J.~L. Sch\"{o}nberger and J.-M. Frahm, ``Structure-from-motion revisited,'' in
  \emph{Conference on Computer Vision and Pattern Recognition (CVPR)}, 2016.

\bibitem{schoenberger2016mvs}
J.~L. Sch\"{o}nberger, E.~Zheng, M.~Pollefeys, and J.-M. Frahm, ``Pixelwise
  view selection for unstructured multi-view stereo,'' in \emph{European
  Conference on Computer Vision (ECCV)}, 2016.

\bibitem{fTetWild}
\BIBentryALTinterwordspacing
Y.~Hu, T.~Schneider, B.~Wang, D.~Zorin, and D.~Panozzo, ``Fast tetrahedral
  meshing in the wild,'' \emph{ACM Trans. Graph.}, vol.~39, no.~4, Jul. 2020.
  [Online]. Available: \url{https://doi.org/10.1145/3386569.3392385}
\BIBentrySTDinterwordspacing

\bibitem{TetGen}
\BIBentryALTinterwordspacing
H.~Si, ``Tetgen, a delaunay-based quality tetrahedral mesh generator,''
  \emph{ACM Trans. Math. Softw.}, vol.~41, no.~2, Feb. 2015. [Online].
  Available: \url{https://doi.org/10.1145/2629697}
\BIBentrySTDinterwordspacing

\bibitem{FuchsiaRadixSort}
\BIBentryALTinterwordspacing
{Google}. (2024) {RadixSort/VK}. [Online]. Available:
  \url{https://fuchsia.googlesource.com/fuchsia/+/refs/heads/main/src/graphics/lib/compute/radix_sort/}
\BIBentrySTDinterwordspacing

\bibitem{FragmentShaderInterlock}
\BIBentryALTinterwordspacing
S.~Grajewski. (2015) {ARB\_fragment\_shader\_interlock}. [Online]. Available:
  \url{https://registry.khronos.org/OpenGL/extensions/ARB/ARB_fragment_shader_interlock.txt}
\BIBentrySTDinterwordspacing

\end{thebibliography}

\end{document}